\documentclass[sigconf]{acmart}

\usepackage{graphicx}
\usepackage{textcomp}
\usepackage{xcolor}
\usepackage{microtype}
\usepackage{cleveref}
\usepackage{booktabs}
\usepackage{xspace}
\usepackage{subcaption}
\usepackage{multirow}
\usepackage{array}
\usepackage{balance}
\usepackage{tikz}

\newcommand{\eg}{e.g.,\xspace}

\definecolor{atomgray}{RGB}{110, 110, 110}
\definecolor{listblue}{RGB}{37, 99, 235}
\definecolor{tablegreen}{RGB}{22, 163, 74}
\definecolor{blockpurple}{RGB}{152, 16, 250}

\newcommand{\structname}[2]{%
  \textcolor{#1}{#2}%
}

\newcommand{\styleatom}[1]{\structname{atomgray}{#1}}
\newcommand{\stylelist}[1]{\structname{listblue}{#1}}
\newcommand{\styletable}[1]{\structname{tablegreen}{#1}}
\newcommand{\styleblock}[1]{\structname{blockpurple}{#1}}

\newcommand{\sysatom}{Atom\xspace} % not yet used everywhere, bc of ambiguity of cell vs label
\newcommand{\sysatoms}{Atoms\xspace} % not yet used everywhere, bc of ambiguity of cell vs label
\newcommand{\sysAtom}{Atom\xspace} % not yet used everywhere, bc of ambiguity of cell vs label
\newcommand{\sysAtoms}{Atoms\xspace} % not yet used everywhere, bc of ambiguity of cell vs label
\newcommand{\syscell}{Atom\xspace} % not yet used everywhere, bc of ambiguity of cell vs label
\newcommand{\syscells}{Atoms\xspace} % not yet used everywhere, bc of ambiguity of cell vs label
\newcommand{\sysCells}{Atoms\xspace} % not yet used everywhere, bc of ambiguity of cell vs label

\newcommand{\systable}{Table\xspace}
\newcommand{\systables}{Tables\xspace}
\newcommand{\sysTable}{Table\xspace}
\newcommand{\sysTables}{Tables\xspace}

\newcommand{\syslist}{List\xspace}
\newcommand{\syslists}{Lists\xspace}
\newcommand{\sysList}{List\xspace}
\newcommand{\sysLists}{Lists\xspace}

\newcommand{\sysblock}{Block\xspace}
\newcommand{\sysblocks}{Blocks\xspace}
\newcommand{\sysBlock}{Block\xspace}
\newcommand{\sysBlocks}{Blocks\xspace}

\newcommand{\syslabel}{Label\xspace}
\newcommand{\syslabels}{Labels\xspace}

\newcommand{\sysLabels}{Labels\xspace}

\newcommand{\systemplate}{template\xspace}

\newcommand{\sysTemplates}{Templates\xspace}

\newcommand{\dpone}{\textbf{DP1}\xspace}
\newcommand{\dptwo}{\textbf{DP2}\xspace}

\newcommand{\usertext}[1]{\textsf{``#1''}}
\newcommand{\gridloc}[1]{\textit{#1}} % how to style A1 etc
\newcommand{\systemtext}[1]{{\small\texttt{#1}}} % how to style text typed into the system

\definecolor{lightgray}{RGB}{240, 240, 240} 

\newcommand{\formulatext}[1]{{\setlength{\fboxsep}{0.1pt}\colorbox{lightgray}{\small\texttt{#1}}}} % how to style formulas

\definecolor{celery}{RGB}{81, 157, 62}

\newcommand{\circleletter}[2][celery]{%
  \tikz[baseline=(char.base)]{%
    \node[
      circle,
      fill=#1,
      draw=white,
      line width=0.5pt,
      minimum size=0.1em,
      inner sep=0pt,
      outer sep=0pt
    ] (char) {%
      \textsf{\textbf{\color{white}\makebox[0.62em][c]{\raisebox{0.1em}[0.62em][0.0em]{\scalebox{0.65}{#2}}}}}%
    };%
  }%
}
\newcommand{\circleletternexttoparen}[2][celery]{%
  \raisebox{-0.05em}{\circleletter[#1]{#2}}%
}

\copyrightyear{2026}
\acmYear{2026}
\setcopyright{cc}
\setcctype{by}
\acmConference[UIST '26]{The 39th Annual ACM Symposium on User Interface Software and Technology}{November 02--05, 2026}{Detroit, MI, USA}
\acmBooktitle{The 39th Annual ACM Symposium on User Interface Software and Technology (UIST '26), November 02--05, 2026, Detroit, MI, USA}
\acmDOI{10.1145/3830398.3830625}
\acmISBN{979-8-4007-2856-3/2026/11}

\begin{document}

\title{Cellery: Gradual, Composable Structure for Spreadsheets}


\author{Jacob Yim}
\affiliation{
  \institution{University of California San Diego}
  \city{La Jolla, California}
  \country{USA}}
\email{j1yim@ucsd.edu}

\author{Sathwika Peechara}
\affiliation{
  \institution{University of California San Diego}
  \city{La Jolla, California}
  \country{USA}}
\email{speechara@ucsd.edu}

\author{Brian Hempel}
\affiliation{
  \institution{University of California San Diego}
  \city{La Jolla, California}
  \country{USA}}
\email{bhempel@ucsd.edu}

\author{Michael Coblenz}
\affiliation{
  \institution{University of California San Diego}
  \city{La Jolla, California}
  \country{USA}}
\email{mcoblenz@ucsd.edu}


\begin{abstract}
Structure in spreadsheets is typically restricted to ranges of cells and flat tables. However, replacing the open-ended two-dimensional grid with a more structured alternative risks limiting users' freedom and impeding exploratory workflows. How can we enable expressive structure in spreadsheets while preserving their flexibility? We propose a model that allows users to gradually add structure on top of the traditional grid. To identify common structural patterns, we conducted a content analysis of 30 real-world spreadsheets. Based on our findings, we developed Cellery, a novel spreadsheet system that makes these structures explicit. In Cellery, users gradually define the structure of their spreadsheets using named groupings of cells, composing together four structural primitives: Atoms, Lists, Tables, and Blocks. Once defined, users can directly manipulate structures, dynamically reference them in formulas, and incorporate them in reusable abstractions. We evaluate Cellery with an expressiveness study reproducing spreadsheets from our sample, demonstrating that it is capable of representing types of structure that traditional spreadsheets cannot. In a preliminary user study of five participants, we find that they successfully use Cellery to design spreadsheets with explicit structure, and we uncover opportunities for further work on the formula language.
\end{abstract}




\begin{teaserfigure}
  \includegraphics[width=\textwidth]{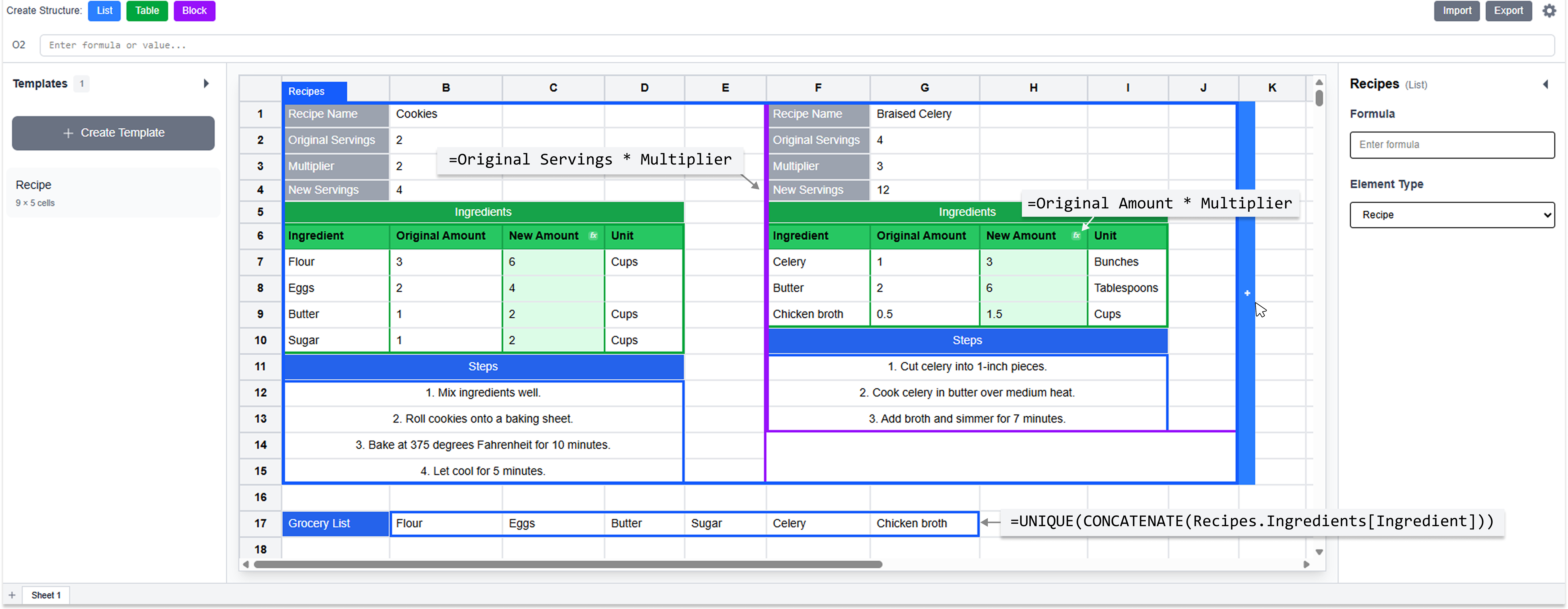}
  \caption{In Cellery, spreadsheets can be structured by combining four structure types: \styleatom{\sysatoms}, \stylelist{\syslists}, \styletable{\systables}, and \styleblock{\sysblocks}.}
  %, indicated by colored borders on the grid. 
  % This example displays a spreadsheet storing a List of recipes. Each recipe has a Table of ingredients and a List of steps. Using a formula, this spreadsheet applies a multiplier to the number of servings to compute the amount of each ingredient needed. 
  %Cellery is designed to support expressive nested structure: the spreadsheet in \Cref{teaser} consists of a List of Blocks, each containing a Table, a List, and a number of Atoms.
  \label{teaser}
  \Description[The Cellery user interface.]{Screenshot of the Cellery spreadsheet user interface showing a collection of recipes organized as regions on a grid of cells. Structural elements are visually indicated with colored outlines, displaying Lists in blue, Tables in green, and Blocks in purple. The spreadsheet contains a list of recipe blocks arranged horizontally, with two recipes: one for Cookies and one for Braised Celery. Each recipe is represented as a block containing multiple nested structures: labeled single-value fields including recipe name, original servings, multiplier, and new servings; a table of ingredients with columns for ingredient name, original amount, new amount, and unit; and a list of preparation steps. Below the list of recipes, a grocery list stores the unique ingredients needed to make both recipes. Text boxes display the formula for computing the New Servings (=Original Servings * Multiplier), New Amount (=Original Amount * Multiplier), and Grocery List (=UNIQUE(CONCATENATE(Recipes.Ingredients[Ingredient]))). The interface has a toolbar at the top with buttons for creating structures, and two panels: on the right is a panel for inspecting structures (e.g., displaying their formula), and on the left is a panel containing template definitions (including the template defining a Recipe).}
\end{teaserfigure}

% cream of celery
% ants on a log
% nutella french toast
% 

\maketitle

\section{Introduction}

% 1. Spreadsheets derive their power from a flexible model
% 2. But that model is limited to cells and flat tables
% 3. As a result formulas are fragile + have limited expressivity, manual work involved in structural reuse
% 4. Prior work tried to close this gap by imposing structure on users, but did so by restricting the flexible model that makes spreadsheets so valuable in the first place
% 5. However, we hypothesize that spreadsheets *already* carry implicit structure: it doesn't need to be enforced on users because it already exists, it just isn't made explicit
% 6. To understand what types of implicit structure people create in spreadsheets we ran our formative content analysis, and based on the findings we develop Cellery.

Spreadsheets, used by tens of millions of people~\cite{scaffidi_estimating_2005}, are ubiquitous for working with data. The spreadsheet's open-ended two-dimensional grid of cells enables users' exploratory workflows and preserves their sense of agency~\cite{chalhoub_its_2022}. The freeform grid lets users add data without needing to specify any structure upfront.
Through direct manipulation and instant visual feedback, even inexperienced users without deep knowledge of spreadsheet features (e.g., the formula language) can quickly begin working with the system and gently progress to more advanced skills as needed~\cite{nardi_small_1993}.

Many tasks require working with rich structures like nested groups, hierarchies, or heterogeneously-sized collections: all types of data that may be difficult to fit into a flat, two-dimensional form. Unfortunately, spreadsheet users must choose between two unsatisfactory ways of representing these structures. They can impose visual structure with styling, such as bold cell borders; however, these informal representations of structure are invisible to the system, and not usable in contexts like formulas. Alternatively, they can flatten their data, but these transformations may be difficult and complicate analysis because they impact visual layout.

% as might be used in a database

% Despite these benefits, documents often in fact bear richer structure, and when the limited flat table features do not suffice, users may instead impose visual structure with style choices (such as bold cell borders). Unfortunately, these style choices do not create \textit{semantic} structure: the structure is invisible to the system. This is a common scenario, since many tasks require working with richer structures like nested groups, hierarchies, or heterogeneously-sized collections: all types of data that may be difficult to fit into a flat, two-dimensional form. Users must either flatten their data to satisfy the model, or give up formal representation and communicate structure using more informal means, like visual conventions.

This lack of expressivity has consequences for users. First, since the data model of a spreadsheet is a flat grid, formulas must be written in terms of grid coordinates instead of domain structure. Even with features like named columns in tables, formulas using other types of structure, like hierarchies and nested groupings, may be difficult to write without hard-coded references to cell addresses. Formulas of this kind are fragile: structural change (for example, addition of new data) may cause cell addresses to shift and formulas to break. This can lead to errors and increases the cost of spreadsheet maintenance and reorganization, since formulas must be edited whenever structure changes. Structural \textit{reuse} presents another limitation: when a structural pattern needs to be replicated, it must be copied and edited manually, with the system blind to relationships between instances. Finally, user \textit{interactions} with structure are limited to what the system can represent: beyond flat tables and ranges, users cannot directly select or operate on structure.

As a result, there have been several efforts over the years to introduce more structured approaches to spreadsheets. Prior systems have committed to constrained structural models, such as hierarchies~\cite{chang_using_2016, mccutchen_object_2016}, object-oriented classes~\cite{engels_classsheets_2005}, or relational schemas~\cite{bakke_spreadsheet-based_2011}. While these tools may be especially valuable for users working with specific types of data, they also replace the general-purpose, flexible grid that is so vital to the spreadsheet's success. Spreadsheet work is frequently exploratory and unplanned~\cite{brown_experimental_1987, panko_what_1998}: users may begin a task without knowing what the structure of their data will look like, developing their understanding through loosely-structured experimentation.
%Users may also work with varied, heterogeneously structured data that does not fit cleanly into a single paradigm.
Replacing the open-ended, two-dimensional grid may be too significant a loss of flexibility for many users~\cite{chalhoub_its_2022}.

% We propose a model of \textit{gradual} structure~\cite{miller_gradual_2016} that builds on top of the cell-based model instead of replacing it.
% how can we can design a system that allows users to freely structure their spreadsheets in expressive ways without requiring them to fit their data to a specific format?
% how can we can design a system that allows users to freely structure their spreadsheets in expressive ways without requiring them to commit to a specific format upfront?
% MC suggestion: How can a spreadsheet system gradually bridge the gap between unconstrained grids and expressive structure?
% how can a spreadsheet system support both expressive structure and user freedom?

Our research question is: \textit{how can spreadsheets support expressive structure without sacrificing their flexibility?} Between the unconstrained grid and prescriptive, predefined structure, is there a middle ground? One answer may lie in the ways users currently structure their spreadsheets. We hypothesize that spreadsheets \textit{already} carry rich structure, but that it is left \textit{implicit} and invisible to the system. We propose a model of \textit{gradual} structure that builds on top of the flexible two-dimensional grid instead of replacing it, allowing users to incrementally add structure on top of existing data.

% To understand existing implicit structure in spreadsheets and identify what types of structure users create in practice

To understand the types of implicit structure spreadsheet users currently create, we first conducted a content analysis of 30 real-world spreadsheets from the EUSES corpus~\cite{fisher_euses_2005}. We identified common organizational patterns and formulated design opportunities for improving the expressive power of structure in spreadsheets.

% Cellery does not replace the grid with a schema.
% Cellery makes structure a manipulable, spatial object.
% Based on our findings, we developed a novel spreadsheet system, Cellery. In Cellery, users can gradually define the structure of their spreadsheets by composing together four primitive structure types: Atoms, Lists, Tables, and Blocks.

% \jy{Typical spreadsheet systems have a data model consisting of a grid, and cells can contain data. However, this limits users' ability to reference structure because formulas must reference geometry, even though the geometry can change. In our system, formulas reference structures. Cellery is a grid with objects on it.}

These observations informed our design of \emph{Cellery}, a novel spreadsheet system that allows users to gradually define structure directly on the grid. Rather than a flat grid of values addressed by location, Cellery's data model consists of named structures that reside on the grid. Users compose together four primitive structure types: \sysatoms{}, \syslists{}, \systables{}, and \sysblocks{}.\ \sysAtoms{} hold single values and are created when users edit a grid location; \syslists{} are one-dimensional, ordered collections of structures; \systables{} are two-dimensional grids of structures; and \sysblocks{} are groupings of structures without ordering or tabular organization. 
%Composability makes the internal organization of these structures explicit:
%Formulas can reference these structures by name so that references remain stable when structures move.
Once defined, these structures provide various capabilities: they can be referenced by name in formulas,
%(making references stable when structures move and change)
directly manipulated to perform operations like expansion, and incorporated into reusable abstractions over structural patterns.

\Cref{teaser} shows a spreadsheet in Cellery consisting of a collection of recipes, each with a table of ingredients and list of instructions. A multiplier on each recipe changes the number of servings produced and ingredients needed. Cellery supports nested, heterogeneous structure: this spreadsheet consists of a \syslist{} of \sysblocks{}, each containing a \systable{}, a \syslist{}, and several \sysatoms{}. Structures can be given names, which can be placed on the grid as \syslabel{} \sysatoms{} (as shown in \systemtext{Recipe Name}, \systemtext{Steps}, etc.). Formulas use these names to reference structures: a \mbox{\systable{}} column formula, \formulatext{=Original Amount * Multiplier}, references the \syscell{} labeled \systemtext{Multiplier} to compute the count of each ingredient. A construct called a \textit{\systemplate} defines a reusable structural pattern: this spreadsheet incorporates a custom \systemtext{Recipe} template, so pressing a single ``+'' button automatically expands the \syslist{} and creates a new \systemtext{Recipe} instance. %\jy{TODO: put callout numbers on the figure to reference them in this description.}

Through an expressiveness study using spreadsheets from our corpus, we demonstrate that Cellery can explicitly represent forms of structure that remain implicit in traditional spreadsheets.
%including nested structure, hierarchies, and repeated structural patterns.
Finally, with a preliminary user study of five participants, we find that participants use Cellery to design spreadsheets with rich structure, but may struggle to leverage that structure using formulas.

We focus this work on the design of Cellery's structural model and its interface.
%since explicit structure is a prerequisite allowing for richer spreadsheet formulas and interactions.
Cellery's formula language demonstrates what structure could enable; we leave more focused design and evaluation of the language to future work (\Cref{discussion}).

This paper contributes:
%\jy{Center composable, empirically grounded structural model as our main contribution? One framing I haven't really explored is that graduality isn't just a feature of the system, it's also a feature of the model, because our structures layer cleanly on top of the flat grid}
\begin{enumerate}
    \item A content analysis of 30 real-world spreadsheets, yielding structural patterns that motivate our design.
    \item A model of composable structure for spreadsheets, featuring four primitives (\sysatoms, \syslists, \systables, and \sysblocks) that together form expressive structure on the grid.
    \item A novel spreadsheet system, Cellery, implementing our composable structure model paired with structure-aware formulas, interactions, and custom abstractions.
    \item An expressiveness evaluation reproducing spreadsheets from our content analysis, alongside a preliminary user study of the system (n=5).
\end{enumerate}

% \begin{enumerate}
%     \item A content analysis of 30 real-world spreadsheets, yielding a taxonomy of structural patterns and five design opportunities for systems.
%     \item A novel spreadsheet system, Cellery, featuring a model of composable structures paired with structure-aware formulas, interactions, and custom abstractions.
%     \item An expressiveness evaluation demonstrating that Cellery can represent the structural patterns identified in our corpus, and characterizing the limits of our model.
%     \item A think-aloud study examining how 6 participants use Cellery for tasks involving spreadsheet authoring and reuse.
% \end{enumerate}

% \mc{Get a figure showing the main ideas of Cellery into the intro (because this is primarily a systems paper, not an empirical study paper). Walk the reader through the main features.}

% \jy{Reduce amount of related work in the intro, describe more of the system here, and move the related work next}

\section{Related Work}
\label{relwork}
We discuss related work in three categories: approaches for supporting structure within traditional spreadsheet systems; alternatives to the two-dimensional grid; and empirical studies of spreadsheets.
% Table of systems and features

\paragraph{Traditional spreadsheets.}
% Excel tables, make sure to compare against pivot tables. Named ranges
Popular spreadsheet systems like Excel and Google Sheets offer several features for adding structure on top of the two-dimensional grid.\ \textit{Named ranges} allow for cell regions to be referenced in formulas by name rather than by address.\ \textit{Tables} associate column names with data: for instance, the formula \formulatext{=AVERAGE(Table1[Column1])} computes the average of \systemtext{Column1}, regardless of the number of rows.\ \textit{Dynamic array formulas} return multiple values that ``spill'' into adjacent cells.\ \textit{Pivot tables} summarize tabular data into aggregate views.

These features provide helpful structure, but they ultimately still work within the confines of the flat, cell-based model. Most notably, they do not capture the nested, hierarchical structure that Cellery's compositional model provides. Named ranges carry no structural information about the cells inside beyond their location; additionally, they do not possess explicit operations for expanding or adding data. Tables are most similar in approach to Cellery's structures; columns can be referenced by name rather than location, and tables can also be transformed with structure-aware operations (e.g. add columns, move column, move table). However, they are still limited to flat layouts, and cannot directly represent nesting or hierarchy. Dynamic arrays are functionally similar to Cellery formulas that return \syslists{}, but they are unaware of surrounding structure: formulas that spill onto existing values cause errors instead of accommodating them. Finally, pivot tables are powerful for summarization, but they are view-only and require source data to be flat: they are a means of viewing data, not structuring it. 

% One tool for adding structural abstractions to a traditional spreadsheet is Gridlets.

One structural extension to spreadsheets proposed by prior work is the Gridlet~\cite{joharizadeh_gridlets_2020}, an Excel function outputting a live copy of a range of cells with value overrides. Gridlets, similarly to Cellery's templates (\Cref{templates}), serve as reusable abstractions over spreadsheet regions. The main difference is that Cellery's templates abstract over composable structures rather than flat ranges of cells, enabling variable-length substructures in templates. Additionally, Cellery supports abstract template definitions with configurable defaults, rather than requiring references to existing instances.

Inferring structure in spreadsheets has been useful for identifying errors in formula logic and references~\cite{abraham_header_2004, abraham_inferring_2006, abraham_ucheck_2007, zhang_semantic_2021, mittermeir_finding_2002, koci_machine_2016, koci_table_2017}. Prior studies show that certain types of structure are common: one study identified 3,290 tables with headers in 3,110 spreadsheets~\cite{zhang_semantic_2021}. These works demonstrate that spreadsheets carry valuable latent structure; Cellery extends this idea by allowing users to explicitly define new forms of structure, rather than inferring them.

%\jy{Other systems support hierarchies, but they do so by modifying the grid instead of allowing for hierarchies within the existing model.}

\paragraph{Alternatives to the two-dimensional grid.} Several research systems have addressed the limitations of the flat grid by introducing richer structural models~\cite{mccutchen_object_2016, bakke_expressive_2016, bakke_spreadsheet-based_2011, chang_using_2016, burnett_forms3_2001, chang_spreadsheet_2014, miller_gradual_2016, miller_gradual_2016-2, miller_sumwise_2010}. Espalier, also known as Object Spreadsheets~\cite{mccutchen_object_2016}, replaces the grid with hierarchical collections of typed objects, with columns in the spreadsheet representing attributes and nested groups of columns representing objects. Related Worksheets and its successors SIEUFERD and Ultorg~\cite{bakke_expressive_2016, bakke_spreadsheet-based_2011} use the spreadsheet as an interface for a relational database, representing one-to-many and many-to-many relationships via cells holding multiple values, alongside a formula language for constructing SQL-like queries. Gneiss~\cite{chang_spreadsheet_2014, chang_using_2016} makes externally-structured hierarchical data (e.g., JSON retrieved from web APIs) navigable and manipulable within a spreadsheet interface. The spreadsheet grid itself adopts a hierarchical structure, with cells nested inside grouped rows.
%The Lish data model~\cite{hall_structuring_2018} proposes replacing the grid with cells organized into nested lists, to which the user may optionally apply templates for prototyping repeating structures.
The key difference between Cellery and these existing systems is that they all commit to a particular organizational paradigm, whether object-oriented (Espalier), relational (SIEUFERD), or hierarchical (Gneiss). 
%or nested list-based (Lish).
These tools provide deep support for specific types of structure; while this may be useful in specific domains, specialized structure can limit users who do not yet know the organization of their spreadsheet, experiment with different approaches, or work with different types of structure simultaneously.
In contrast, structure in Cellery is composable, heterogeneous, and optional: our four primitives can be freely combined to represent diverse organizational patterns, allowing users to define their own structure rather than adopt a paradigm prescribed by the tool. Additionally, some systems, like Espalier, Forms/3~\cite{burnett_forms3_2001}, Sumwise~\cite{miller_sumwise_2010}, and the Lish data model~\cite{hall_structuring_2018}, fully replace the two-dimensional grid with a more structured or restricted alternative, which may limit user freedom and introduce premature commitment~\cite{blackwell_cognitive_2001} when structure must be specified up front. Gradual Structuring~\cite{miller_gradual_2016}, which helped inspire Cellery's gradual approach (\Cref{gradual}), begins as a flat grid but still introduces alterations (e.g., hierarchical rows/columns) when structure is added. Cellery, in contrast to these systems, builds structure directly \textit{on top} of the familiar grid rather than replacing it.

% Maybe separate object spreadsheets and lish into a separate grid replacement section, sieuferd and gneiss are more like domain-specific spreadsheets

% \subsubsection{Domain-specific spreadsheets}
% Gneiss; SIEUFERD/Related Worksheets/Ultorg, Airtable, Grist

% \subsubsection{Spreadsheet modeling}
% ClassSheets, MDSheet, other Erwig/Cunha, Model Master, Isakowitz

Another direction is model-driven spreadsheet development, which involves generating provably correct spreadsheets from specifications~\cite{engels_classsheets_2005, erwig_gencel_2006, cunha_mdsheet_2012, cunha_bidirectional_2012, cunha_model-based_2014, paine_ensuring_2008, abraham_visual_2005, erwig_automatic_2005}. 
%Gencel~\cite{erwig_automatic_2005, erwig_gencel_2006} defines a formal calculus for preventing errors in spreadsheet formulas; its specification language ViTSL~\cite{abraham_visual_2005} provides a grid-based visual editor for designing templates from which correct spreadsheets are generated.
For instance, ClassSheets~\cite{engels_classsheets_2005}
% built on the Gencel~\cite{erwig_automatic_2005, erwig_gencel_2006} formal calculus and the ViTSL~\cite{abraham_visual_2005} visual editor,
generates spreadsheets from object-oriented, UML-based templates.
%MDSheet~\cite{cunha_mdsheet_2012, cunha_bidirectional_2012} builds on this work further by inferring ClassSheets models from spreadsheets and bidirectionally syncing edits between the sheet and the model.
%The common thread shared by these systems is that there is a separate model specifying the structure and behavior of the spreadsheet
These systems all involve a separate model specifying spreadsheet structure: users must maintain a schema that is divorced from the main grid, and must express structure using that schema's specification language. Cellery's structures, by contrast, live directly on the grid alongside data; there is no separate specification artifact.
%In addition, ViTSL and ClassSheets require structure to be specified up front, limiting the exploratory, gradually-structured processes we aim to support. MDSheet partially addresses this through model co-evolution, but edits to the spreadsheet must still conform to model-defined operations to register as structural changes. 

% \subsubsection{Reusable abstractions in spreadsheets}
% % Gridlets, Interaction Substrates (proposed)

% \subsubsection{Spreadsheet structure inference}
% % a bunch of stuff

% \subsubsection{Studies of spreadsheet users}
% % Chalhoub (most important citation by far), TableTalk, Brown, Struggling to Excel, Nardi?, Dourish?
% % Note: Probably discuss Nardi and Dourish (theoretical foundations) later in system section instead of here, and also Miller

% \subsubsection{Studies of spreadsheet errors}
% % Spreadsheet errors: Panko, Powell

% \subsubsection{Studies of spreadsheet structure}
% Prior empirical work investigating structure in spreadsheets has largely focused its analysis on tables~\cite{zhang_semantic_2021} and logical formula dependencies~\cite{mittermeir_finding_2002}.

\paragraph{Empirical studies of spreadsheets.} In spite of much work on spreadsheet errors \cite{panko_what_1998, panko_spreadsheets_1996, panko_spreadsheet_2008, powell_critical_2008}, few prior works have used an open-ended approach to study \textit{types of structure} in spreadsheets. A manual analysis~\cite{milazzo_emergence_2016} identified tabular patterns in spreadsheets from the EUSES and Enron corpora, finding various types of tables including vertical, single-entry, relationship, hierarchical, and replicated tables. These findings are consistent with ours regarding diversity in tabular structures and repeated structural patterns. Liang et al.~\cite{liang_tabletalk_2025} focused specifically on spreadsheet templates, using an open coding methodology to find that spreadsheet usability is influenced by requirements and structure. They identify two levels of spreadsheet structure, spreadsheet-level and table-level, as well as two types of tables, data tables and insight tables (summaries of the data). Unlike these prior studies, our content analysis examines patterns in the layout of spreadsheets broadly, beyond tabular structures. 
% Notably, our analysis also investigates nested structures, which to our knowledge have not been within the scope of prior empirical research.

%\jy{Reiterate Chalhoub/Sarkar if space}

\begin{table*}[!tb]
\caption{The codebook of structural patterns identified during our content analysis (n = 30 workbook files), sorted by frequency. ``\# Files'' refers to the number of files in our sample containing at least one instance of the corresponding structure.}
\small
\centering
\newcolumntype{L}{>{\raggedright\arraybackslash}p{11cm}}
\begin{tabular}{lrL}
\toprule
\textbf{Category} & \textbf{\# Files (\%)} & \textbf{Description} \\
\midrule
\textbf{Table}                & 29 (97\%) & A rectangular region organized into rows and columns, with rows representing records.\ \\
\textbf{Column headers}       & 29 (97\%) & A row of cells labeling columns.\ \\
\textbf{Label cells}          & 22 (73\%) & A cell whose content names a nearby cell or group (e.g., ``Total:''). Does not include table headers.\ \\
\textbf{Row headers}          & 12 (40\%) & A column of cells labeling rows.\ \\
\textbf{Header hierarchy}     & 12 (40\%) & Headers organized into multiple levels, with grouped subsets of columns or rows.\ \\
\hspace{0.5em} Merged         &  8 (27\%) & Hierarchy indicated by merging adjacent header cells.\ \\
\hspace{0.5em} Non-merged     &  7 (23\%) & Hierarchy indicated by styling alone (e.g., indentation).\ \\
\textbf{1D collection}        & 11 (37\%) & An ordered sequence of cells arranged linearly, without tabular organization.\ \\
\textbf{Non-tabular group}    & 11 (37\%) & A grouping of nearby, semantically related cells whose organization is not tabular.\ \\
\textbf{Nested structure}     & 10 (33\%) & A containment relationship in which one structure is spatially and semantically inside of another.\ \\
\textbf{Repeated pattern}     & 10 (33\%) & The same structural arrangement appearing multiple times.\ \\
\bottomrule
\end{tabular}
\label{codebook}
\end{table*}

\section{Content Analysis}
\label{ca}

To design Cellery's foundational structures, we conducted a formative content analysis to discover: \textit{What types of implicit organizational structures appear in real-world spreadsheets?}

\paragraph{Method.} First, we collected 30 spreadsheets by randomly sampling files from the EUSES corpus~\cite{fisher_euses_2005}, a dataset of 4,498 real Excel spreadsheets. % collected primarily through Google search in 2005.
Because we are interested primarily in understanding what types of structures are present, rather than the distribution of counts of structure types, we chose one spreadsheet workbook (i.e., one \texttt{.xls} file, potentially with multiple sheets) as the unit of analysis for our study. During the coding process, for each file, we indicated whether or not each code was present.

The first two authors began by analyzing an initial sample of 10 spreadsheets and discussing to identify a set of preliminary codes.
Using the preliminary codes, the first two authors coded the 10 spreadsheets independently, then compared and discussed until a consensus was reached about the definition of each code. After the pilot sample, the first author annotated the remaining 20 spreadsheets using the agreed-upon codes. We continued to iterate on the codebook throughout, adding new codes as new patterns were observed. After finishing all 30 spreadsheets, the first author then revisited all previously coded spreadsheets to apply changes made to the codebook. Finally, the second author performed a final pass over the codes. Since we coded collaboratively rather than independently, we do not report inter-rater reliability. %\mc{cite McDonald?}

\begin{figure}[!tb]
    \centering
    \includegraphics[width=\linewidth]{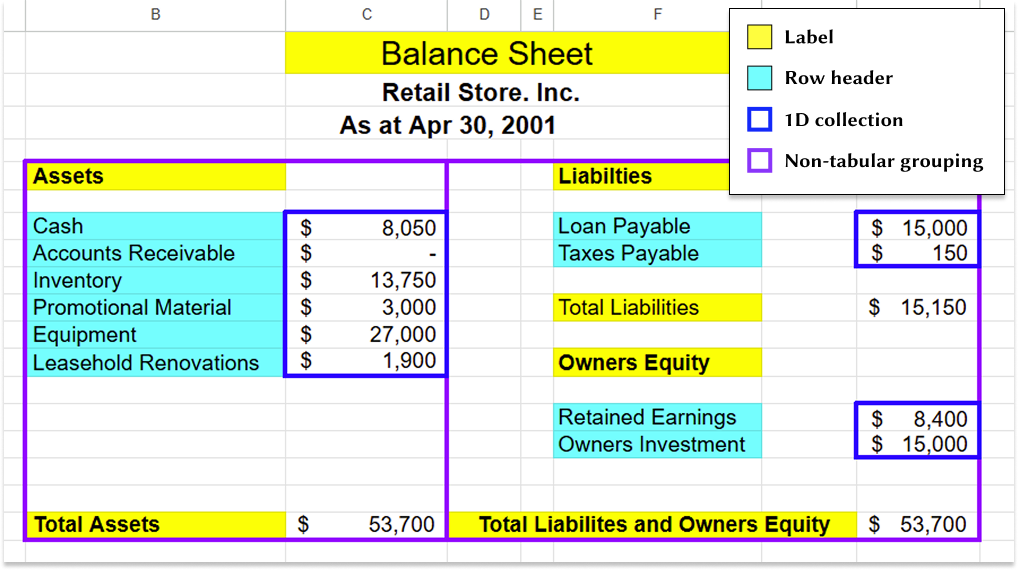}
    \caption{A sampled spreadsheet with labels, 1D collections, non-tabular groupings, row headers, and nested structure.}
    \label{fig:corpus}
    \Description[Example spreadsheet from the EUSES corpus.]{Example spreadsheet from the EUSES corpus illustrating multiple structural patterns. The spreadsheet shows a balance sheet for a retail store, with sections of the spreadsheet recording finances for assets, liabilities, and owners' equity. Cells compute the total assets, total liabilities, and total liabilities and owners' equity. The spreadsheet is marked with color-coded cell backgrounds and borders corresponding to categories from the codebook produced during content analysis: a legend shows labels colored in yellow, row headers colored in cyan, 1D collections surrounded with blue borders, and non-tabular groupings surrounded with purple borders.}
\end{figure}

\paragraph{Results.} \Cref{codebook} displays our codebook of structures and the number of files that leveraged each; \Cref{fig:corpus} displays an example from our corpus. We identified some codes using visual layout: tables, headers, and merged header hierarchies were largely visually unambiguous. However, other codes required interpreting domain meaning: for instance, distinguishing labels from data, or non-tabular groupings from incidentally adjacent data. No new codes emerged starting after coding about 20 spreadsheets, suggesting that our sample size of 30 spreadsheets reached saturation.

\paragraph{Limitations.} We observed a small random sample of 30 spreadsheets (out of a corpus of 4,498 spreadsheets), and as such our findings may not generalize to the whole corpus or other unseen spreadsheets. The EUSES corpus was constructed in 2005 and collected largely via Google search; our analysis may be influenced by the search terms used during construction of the corpus and biased towards older Excel spreadsheets. For instance, Excel tables, introduced in 2003 as Lists, were relatively new at the time of the corpus's construction and may have led to more standardized table layouts in future years.
Our derived codes are largely based on layout, rather than other types of structure (e.g., formula references); we do not claim to identify all forms of implicit structure.

%\jy{Figure showing a spreadsheet with many different codes to get an idea of what they look like?}

\subsection{Opportunities: Representational Gaps}
\label{gaps}

Based on our findings, we identify types of implicit structure that traditional spreadsheet systems cannot represent explicitly. These are \textit{representational gaps}: organizational patterns that users incorporate but that the flat, grid-based model has no means to encode. Each gap presents a design opportunity for improving the structural expressiveness of new systems. We present five such gaps:

\textbf{Labels on the grid} (73\% of files). Labels were present in majority of the spreadsheets in our sample. Cells with semantic information may be useful for helping readers understand a spreadsheet. However, in traditional spreadsheets, labels and data are treated identically: both are cells containing values. As a result, the labeling relationship cannot be leveraged by the system, e.g., in formulas.

\textbf{Hierarchical table headers} (40\% of files). Traditional spreadsheets support explicit tables with single-level column and row headers, but not header hierarchies. We find that spreadsheets in our sample often incorporate hierarchies indicated by layout and styling: 27\% of spreadsheets incorporate hierarchical tables with column or row groups indicated by merged headers, while 23\% of spreadsheets have tables with hierarchy indicated by styling choices, like indentation or bolding. In both cases, hierarchy is not a formal construct, and is instead communicated visually.

\textbf{Non-tabular groupings of cells} (37\% of files). Spreadsheets have limited capacity for representing groupings of cells that do not conform to a row-column pattern. For instance, we observed several form-like spreadsheets consisting of structured regions containing heterogeneous sub-structures, such as the spreadsheet displayed in \Cref{fig:corpus}. Named ranges capture a rectangular region but provide no information about its internal organization; the flat grid model cannot represent or navigate unordered, heterogeneous containers.

\textbf{Nested structure} (33\% of files). Traditional spreadsheets provide no explicit way of representing \textit{nested} structures: structured groups of cells inside of each other. For instance, a flat Excel table cannot represent a table where the row-column intersections hold lists of cells, a pattern that we observed twice in our sample.

\textbf{Repeated structural patterns} (33\% of files). When the same structural arrangement appears multiple times, for instance, a series of tables with the same column headers but different data, the relationship between instances is invisible to the system. We observed a variety of patterns in our sample, including structure repeated across tables and the aforementioned non-tabular groupings.

\section{System Design Principles}
\label{design}
%Based on the findings from our formative content analysis, our motivation of making structure explicit without replacing the flexible grid, and our desire for users to gain the full benefits of structure,

% structure should be \textit{composable}, \textit{gradual}, and \textit{first-class}

% \jy{I think that this design goal section might be too much? It's probably too complex. But I like not having to do too much justification in the system design section, and I want to make the things we value very clear. The design space is really big, so I think we should be opinionated and justify those opinions clearly, or something like that. But I think i may have gone overboard. Lmk if i should cut anything}

Following our content analysis, we guided the development of our system, Cellery, with two high-level design principles:
% \mc{do we need something about supporting the kinds of structures we identified but not everything under the sun? Need to justify why we need a new set of structures instead of just using an existing one such as ClassSheets.}

\dpone: Structure should be \textbf{gradually defined}.

\dptwo: Structure should be \textbf{composable}.

% \textbf{DG3:} Structure should be \textbf{first-class} and integrated throughout the system.

These principles work together towards our goal of enabling structural expressiveness without sacrificing flexibility: graduality preserves the flexibility of the two-dimensional grid, while composability enables expressive user-defined structure.
% , and first-class treatment enables \textit{structure-aware functionality} throughout the system--e.g., in the formula language and direct-manipulation interactions.
We define, motivate, and expand on these goals in the following sections.

% \textbf{Gap 1: Nested structures} are compositions of structures by definition. A composable model makes nesting a formal structural property, rather than an informal property indicated by layout or styling.

% \textbf{Gap 2: Labels on the grid} allow for navigation of composed structures, both for the user reading the spreadsheet, and potentially also for the system (e.g., via formula references).

% \textbf{Gap 3: Non-tabular groupings of cells} enable new forms of composition on the grid: specifically, compositions of structures indicated by containment rather than alignment along an axis.

% \textbf{Gap 4: Tables with hierarchical headers} can be viewed as compositions of tables, where columns are partitioned into sub-tables that share the parent's rows. 
% %\footnote{Note that this is not composition of tables in the strictest sense, where row-column intersections in a table contain other tables. Tables with hierarchical headers might be more precisely described as compositions of partitions of tables sharing an axis.}

% \textbf{Gap 5: Repeated structural patterns} can be represented as abstractions over compositions of structures.

\subsection{Gradual Structure}
\label{gradual}
% Majorly inspired by Miller and Hermans, make sure to cite
% Structure should be optional; it should be incrementally defined; it should be possible to define it in either direction, data first or structure first.
% We should preserve the flexible 2d grid, building structure on top of it instead of replacing it.
% Also talk about explicitness > inference

% Nardi?

To preserve the freedom and flexibility of the two-dimensional grid of cells, we aim to build structure on top of, rather than replace, the familiar grid. We leverage the concept of \textit{gradual structuring} proposed by Miller and Hermans~\cite{miller_gradual_2016-2, miller_gradual_2016}, whereby users incrementally add structure to a spreadsheet as needed.
% \footnote{In programming languages, \textit{gradual typing} allows static types to be used only as needed.}
To focus our design, we highlight four specific aspects of gradual structure:

\textbf{Optionality.} Structure should be optional, giving users freedom to experiment without committing to a specific model.

\textbf{Explicit, not inferred.} To preserve user agency, structure should not be inferred or introduced without explicit user intent~\cite{chalhoub_its_2022}. The system should not prescribe a specific structure for a given use case, but should allow the user to pick from different possibilities.

\textbf{Top-down and bottom-up.} To maximize flexibility and prevent \textit{premature commitment}~\cite{blackwell_cognitive_2001}, structure should be specifiable in both a top-down \textit{and} bottom-up manner: users should be able to introduce structure both before and after adding data.
%Users may want to start with structure first and then add data (similar to defining a schema up front), or they may want to add structure to a spreadsheet with existing data.

\textbf{Preserve layout and organization.} Finally, wherever possible, structure should be applicable to data as it already exists on the grid, without having to reorganize it to fit the system. We want to leverage the organizational patterns that people already use, rather than necessitate restructuring to serve a particular model.

\subsection{Composable Structure}
\label{composable}
% Goal of making spreadsheets more expressive -> formative content analysis revealed actionable ways to do it
% Our design opportunities share the common thread of composability.

Our content analysis revealed five opportunities to improve the structural expressiveness of spreadsheet systems. All five of these opportunities synergize with a \textit{model of composable structures}: labels allow for navigation of composed structures; hierarchical headers represent compositions of sub-tables; non-tabular groupings introduce containment-based, potentially heterogeneous composition; nesting is composition by definition; and repeated patterns can be represented as abstractions over compositions of structures. 

% Walkthrough of an example?

\section{Example Use Case}
\label{example}

Guided by these design principles, we developed Cellery, a spreadsheet system with gradual and composable structure.
To introduce Cellery, we describe a fictional scenario leading to the spreadsheet in \Cref{teaser}.
Sam is preparing to cook dinner and wants to build a recipe book to keep track of ingredients to buy and cooking instructions to follow. To start, she records a recipe for cookies.

\begin{figure}[!tb]
  \makebox[\columnwidth][r]{\includegraphics[width=1.025\columnwidth]{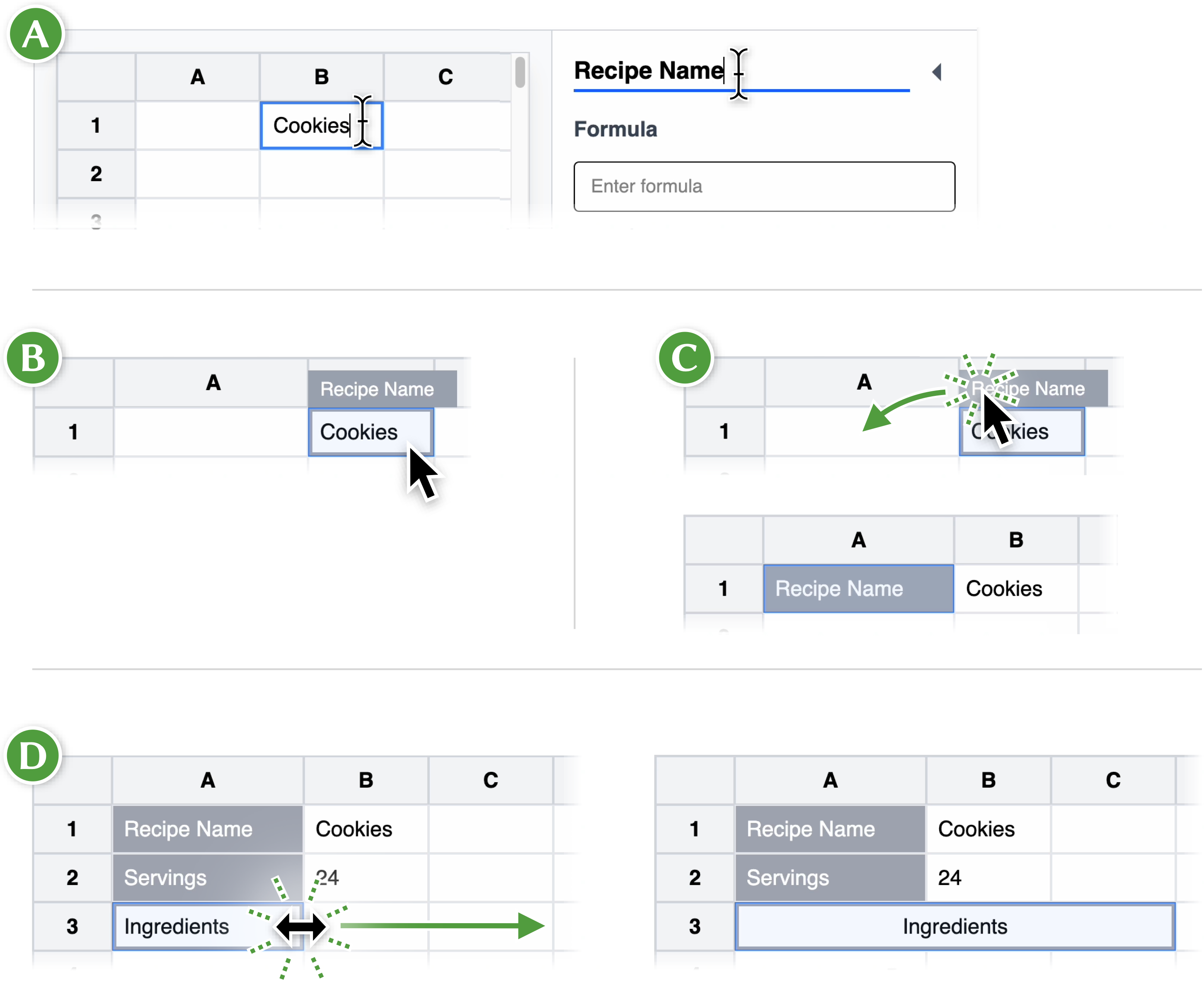}}
  \caption{Sam begins building her recipe book by \mbox{\circleletter{A}\circleletter{B}\circleletter{C}}~creating and labeling \styleatom{\syscells} and \circleletter{D}~resizing structures.}
  \label{fig:example1}
  \Description[A step by step illustration displaying the construction of the structured spreadsheet displayed in Figure 1.]{A step by step illustration displaying the construction of the structured spreadsheet displayed in Figure 1, labeled with letters.
  In part A, a cell with the value "Cookies" is named "Recipe Name" by editing the name value in the structure panel on the right-hand side of the screen.
  In part B, the structure name is displayed as a rectangular gray tab above the cell.
  In part C, the name tab is dragged onto the grid to create a cell containing the name ("Recipe Name"). The cell is a special label cell indicated by gray background.
  In part D, a cell with the value "Ingredients" has its right edge dragged to expand it.}
\end{figure}

\paragraph{\sysCells{} and \sysLabels.}
To title the recipe (\autoref{fig:example1}~\circleletternexttoparen{A}), she double-clicks grid position \gridloc{B1} and types \systemtext{Cookies} to create an \styleatom{\textbf{\syscell{}}}, the basic structure for a single value on the grid (analogous to cells in other spreadsheets). The \textit{structure panel} on the right sidebar lets her name the \syscell{} \systemtext{Recipe Name}.
By default, structure labels appear as a tab~\circleletter{B}~when a structure is selected or when hovering over the structure.
To instead persist the label on the grid, Sam~\circleletter{C}~drags the tab to \gridloc{A1} to reify it as a \syslabel{} \sysatom{}.
\sysLabels can also be created in the reverse order:
% Sam can write text in a \syscell{} first, right-click it, chose \textit{Use as structure label} from the context menu, and click a structure to label.
% , by writing in a grid cell first and then converting the \syscell{} to \syslabel:
Sam writes \systemtext{Servings} and \systemtext{24} in adjacent \syscells{}, right-clicks on \systemtext{Servings}, selects \usertext{Use as structure label} from its context menu (not shown), then clicks the \systemtext{24} \syscell{} to label it.
The link between a \syslabel{} Atom and the named structure is persistent: changing the text in the \syslabel{} Atom's grid location renames the referenced structure, and changing the structure's name in the structure panel changes the value of the linked \syslabel{} Atom.

\paragraph{Resizing.}
When selected, structures can be resized by dragging their edges. For \sysatoms{}, resizing merges cells, allowing Sam to~\circleletter{D}~set up a title for the forthcoming ingredients table.

\begin{figure}[!tb]
  \makebox[\columnwidth][r]{\includegraphics[width=1.025\columnwidth]{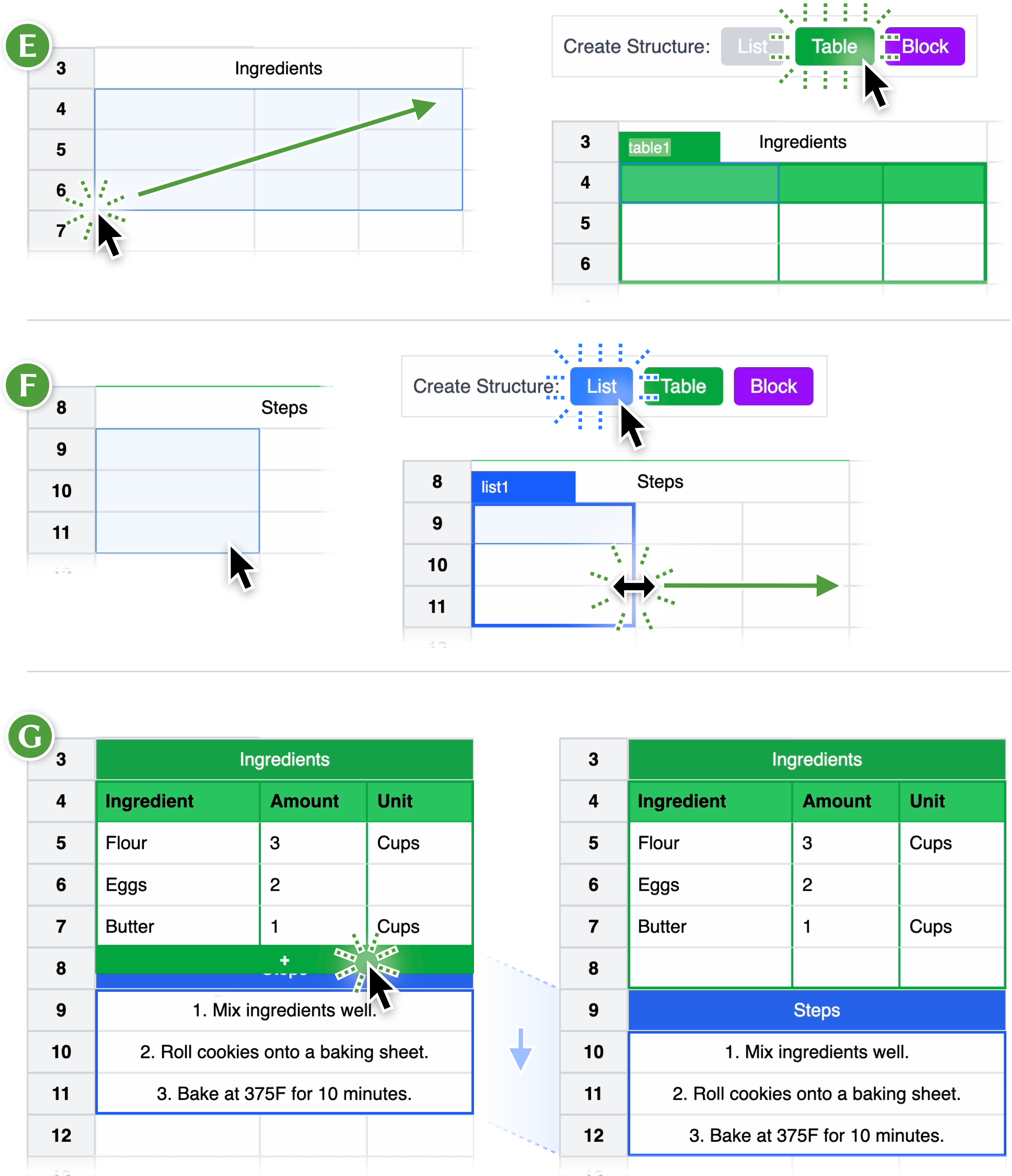}}
  \caption{Sam creates structure with \circleletter{E}~a~\styletable{\systable} and \circleletter{F}~a~\stylelist{\syslist}.\ \circleletter{G}~Expanding a structure repositions other structures.}
  \label{fig:example2}
  \Description[A step by step illustration continuing the sequence from Figure 3 displaying the construction of the spreadsheet in Figure 1.]{A step by step illustration continuing the sequence from Figure 3 displaying the construction of the spreadsheet in Figure 1, labeled with letters.
  In Part E, a 3x3 range of cells is selected by clicking and dragging on the grid, and the "Table" button in the top toolbar is pressed to make the range into a green Table structure. The created Table has 1 row of column headers at the top.
  In Part F, a 3x1 range of cells is selected and the "List" button in the toolbar is pressed to create a blue List structure. The right edge of the List is dragged to make it wider.
  In Part G, a button with a "+" at the bottom of the Table is pressed to add a new row to the table. Doing so pushes down the List below it, indicated by before/after screenshots and a downwards arrow.}
\end{figure}

\paragraph{\sysTables{} and \sysLists{}.}
Because Cellery is gradual, structures can be created first and filled in, or can be applied to existing data. Sam chooses to draw the bounds of her ingredients table ahead-of-time and clicks the ``Table'' button (\autoref{fig:example2}~\circleletternexttoparen{E}), which creates a new \styletable{\textbf{\sysTable{}}} with a header row.
She fills in the \systable{}'s column names and data, and sets the \systemtext{Ingredients} \syscell{} to label the \systable{} (not shown).

She~\circleletter{F}~creates her ingredients list similarly: the ``List'' button initializes a range of cells as a new vertical \stylelist{\textbf{\sysList{}}}, which she resizes to be three columns wide, and fills in with the recipe instructions.

Visibility invites interaction with structures and may help users understand the organization of the spreadsheet. Thus, Cellery always displays structures on the grid---the colored boxes are persistent. This contrasts with named ranges in Excel and Sheets, which must be viewed in a menu or a formula that references the range.

When Sam hovers the cursor over the end of a \systable{} or \syslist{}, a button appears to add another row, column, or item to the structure. Sam~\circleletter{G}~uses this button to make a row for a new ingredient. Cellery automatically maintains the layout of her spreadsheet by moving the \syslist{} of instructions downward to avoid overlap.
In general, Cellery follows two layout rules:
(1) Overlap is disallowed, so \textit{expanding structures push other structures} in the direction of expansion, as in this case. This push is recursive: if pushed structures are moved into other structures, those structures are also pushed.
(2) Parent structures must fully contain their children, so \textit{parents expand with children} if a child expands beyond the parent's bounds.

\begin{figure}[!tb]
  \makebox[\columnwidth][r]{\includegraphics[width=1.025\columnwidth]{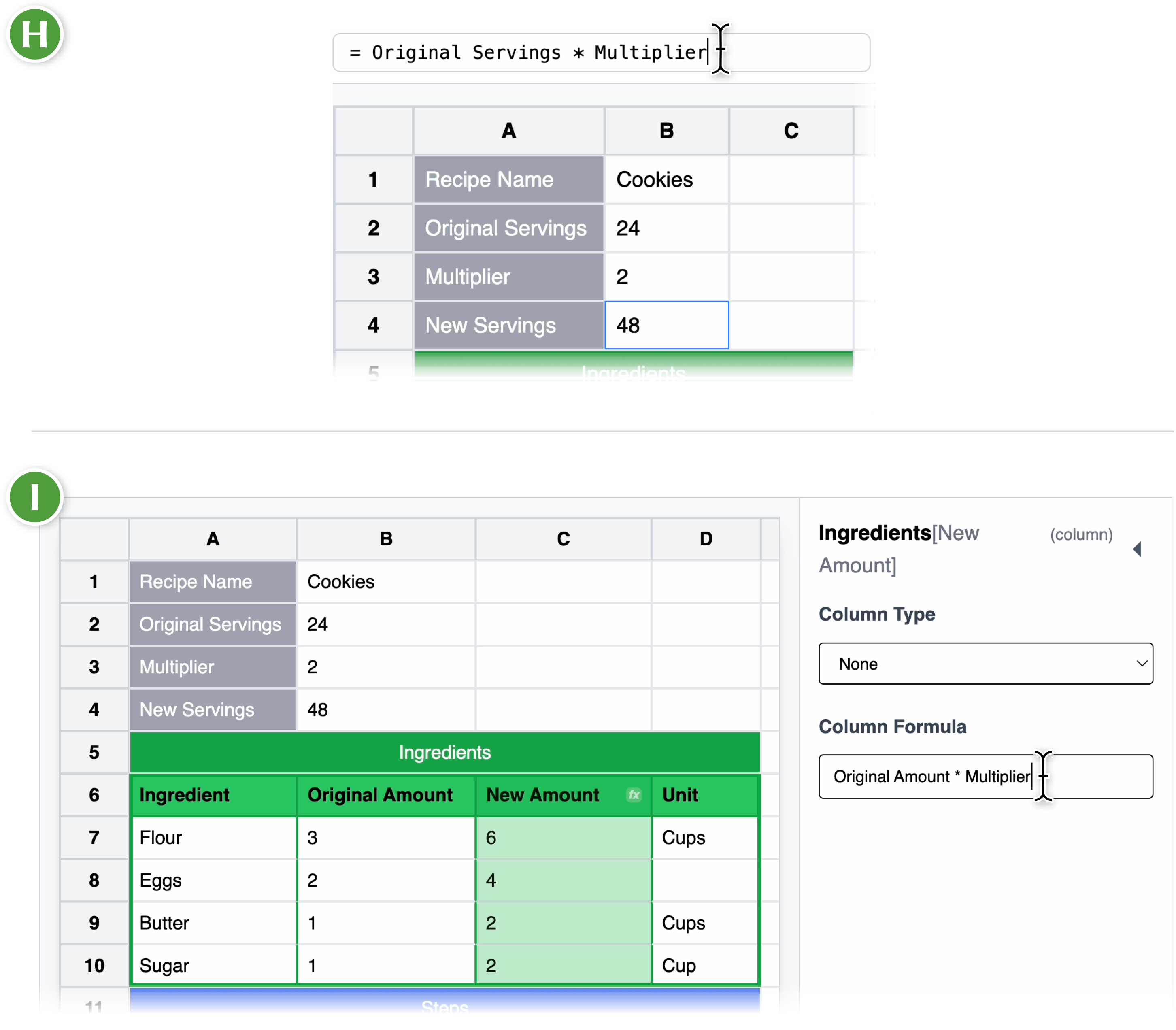}}
  \caption{Sam makes her recipe computational using structural formulas in~\circleletter{H}~an \styleatom{\syscell} and~\circleletter{I}~a \styletable{\systable} column.}
  \label{fig:example3}
  \Description[A step by step illustration continuing the sequence from Figure 4 displaying the construction of the spreadsheet in Figure 1.]{A step by step illustration continuing the sequence from Figure 4 displaying the construction of the spreadsheet in Figure 1, labeled with letters.
  In Part H, a cell labeled "New Servings" is edited using the formula bar at the top of the screen to set its formula to "= Original Servings * Multiplier".
  In Part I, an Ingredients table with columns Ingredient, Original Amount, New Amount, and Unit is edited using the Structure Panel on the right-hand side of the screen to set the "Column Formula" field of the "New Amount" column, setting its value to the formula "Original Amount * Multiplier".}
\end{figure}

\paragraph{Structural Formulas.}
%
% Sam creates \syscells{} storing the recipe name and number of servings, a \systable{} to track ingredients, and a \syslist{} of instructions for preparing the recipe. She uses the tab on top of each structure to name it, and then drags the name tabs onto the grid to create \syslabels{} \textit{labeling} the structures. Expanding the \systable{} to add ingredients automatically maintains the layout of her spreadsheet by moving the \syslist{} of instructions to a larger row index, avoiding overlap.
%
Sam realizes that she needs to make more servings than her recipe provides.
% To make room for a \systemtext{Multiplier} parameter she separately drags the \syslist{}, \systable{}, and their \syslabels{} downward---each individual structure moves as a unit on drag (not shown). She then adds a scale factor \syscell{} labeled \systemtext{Multiplier}.
Labeled structures enable Cellery to support \textit{structural formulas} that reference data by name instead of by grid location.
After adding an \syscell{} labeled \systemtext{Multiplier}, Sam uses the formula \formulatext{=Original~Servings~*~Multiplier} to compute a \systemtext{New Servings} \syscell{}, as shown in \Cref{fig:example3}~\circleletter{H}.
Columns may be computed by formula as well. Sam adds a column using the context menu (not shown) and~\circleletter{I}~uses the structure panel to set a formula for the whole column: \formulatext{Original Amount~*~Multiplier}.

\begin{figure}[!tb]
  \makebox[\columnwidth][r]{\includegraphics[width=1.025\columnwidth]{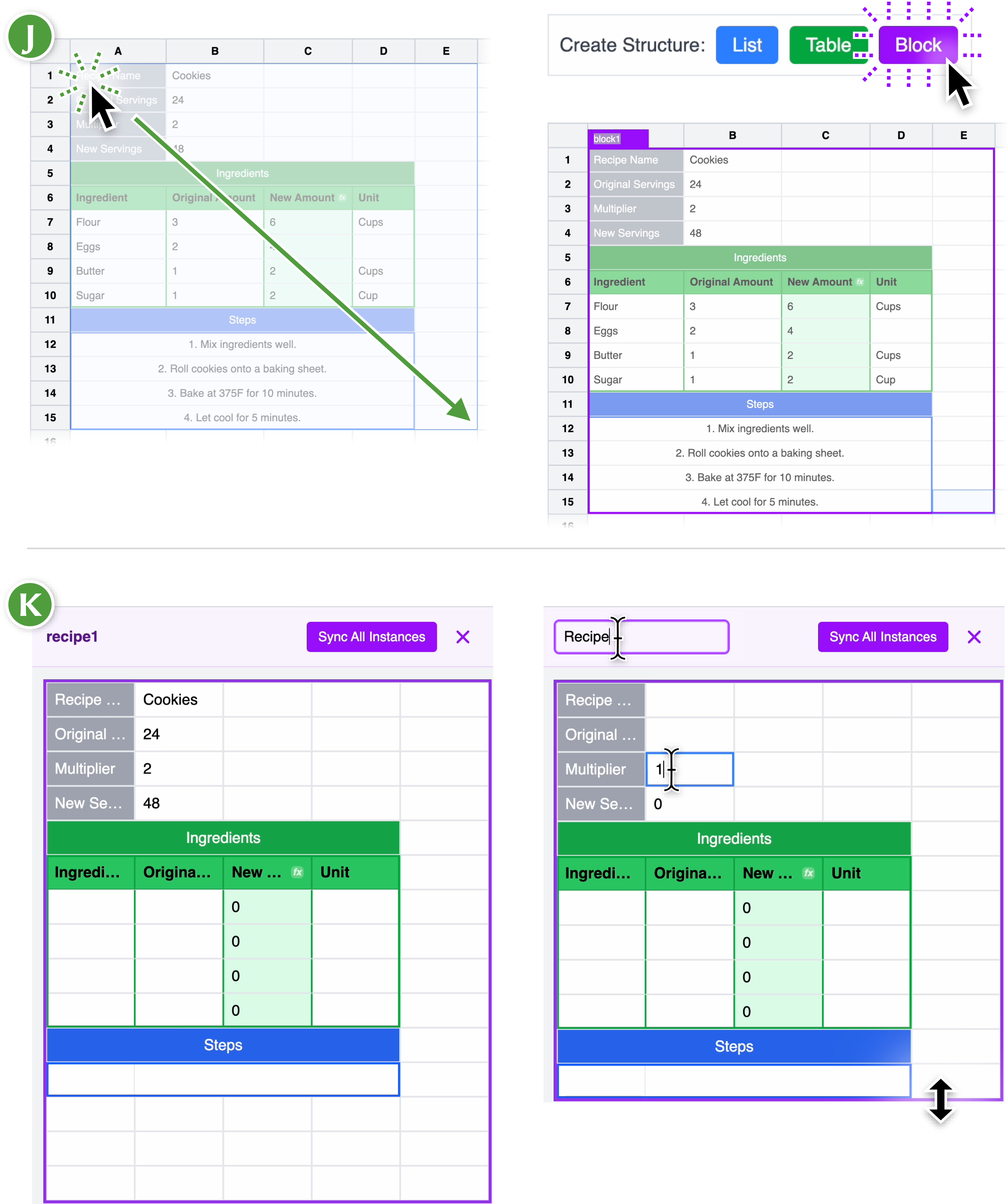}}
  \caption{\circleletter{J} A rectangular \styleblock{\sysblock} serves to group heterogeneous structures into a named object which may be abstracted into a \systemplate, editable with~\circleletter{K}~the template editor.}
  \label{fig:example4}
  \Description[A step by step illustration continuing the sequence from Figure 5 displaying the construction of the spreadsheet in Figure 1.]{A step by step illustration continuing the sequence from Figure 5 displaying the construction of the spreadsheet in Figure 1, labeled with letters.
  In Part J, a range of cells is selected around the structures comprising one recipe (labeled cells for recipe name, servings, and multiplier, table for ingredients, and list of steps). The "Block" button in the toolbar is pressed to create a Block structure.
  In Part K, the template editor for the recipe template is shown. The template has a name at the top of the editor which is edited to "Recipe", and it has a purple border that is resized by dragging.}
\end{figure}

\paragraph{\sysBlocks{} and \sysTemplates.}
Sam decides to add additional recipes to her spreadsheet.
The \styleblock{\textbf{\sysblock{}}} structure in Cellery gathers all the structures within a rectangular region into a single entity. Sam selects a range encompassing the recipe's cells and presses the ``Block'' button (\autoref{fig:example4}~\circleletternexttoparen{J}). This makes all the contained structures children of the new \sysblock{}.
To faithfully communicate structure to users, 
Cellery follows the principle that \textit{spatial containment indicates ownership}.
Moving a child into or out of a parent will reparent it; this applies to moving structures into and out of \syslists{} and \systables{} as well.
Parenting affects referencing: after Sam names the \sysblock~\systemtext{recipe1}, her formulas may reference the structures inside: 
\eg~\formulatext{=recipe1.Recipe Name}.

To reuse \sysblocks{}, Cellery lets users turn a \sysblock{} into \systemplate, for instance, to represent a blank recipe.\ \sysTemplates provide both a default object for new items when expanding a collection (\eg a \syslist) as well as a way to bulk edit all instances.
The right-click context menu on the \systemtext{recipe1} \sysblock{} lets Sam initialize a new \systemplate, which can be edited by selecting it from the \textit{template panel} in the left sidebar (visible in \Cref{teaser}) to open the \textit{template editor}~\circleletter{K}.
% She first creates another structure, a \sysblock{}, grouping together the structures that comprise a single recipe.
% Then, she uses the context menu to convert the \sysblock{} into a reusable \systemplate. Doing so opens a \emph{template editor}.
Inside the \systemplate definition, the \systemtext{Ingredients} \systable{} and \systemtext{Steps} \syslist{} have been cleared automatically. She clears the other fields, names the \systemplate{} \systemtext{Recipe}, and shrinks the \systemplate bounds; instances will later expand to encompass their child structures.
% Sam edits the \systemplate definition to set default values and locations for the structures within, and names the \systemplate \systemtext{Recipe}. Afterward, she creates a new \syslist{} named \systemtext{Recipes}, and drags the \sysblock{} with the recipe for cookies into the \syslist{}.
%\bh{she doesn't drag the list bounds over the recipe? feels more natural here}

Now, to create a list with multiple recipes, back in the spreadsheet Sam makes a new \syslist{}, names it \systemtext{Recipes}, and (not shown) drags the existing \systemtext{recipe1} \sysblock{} into the first cell of the \syslist: this configures the \syslist{} so its elements follow the \sysblock's template, namely \systemtext{Recipe}. Now, when Sam hovers over the end of the \syslist{} and presses the ``+'' button, a new \systemtext{Recipe} instance is automatically created. Sam can now add more recipes without re-specifying their structure, and their content is individually editable. She fills in a recipe for Braised Celery (shown in \Cref{teaser}).
At this point, the structures are quite deep: \eg a \systemtext{recipes} \syslist{} containing a \systemtext{Recipe} \sysblock{} containing a \systemtext{Steps} \syslist{} containing an \syscell{}.
For focus in the structure panel, when multiple structures occupy the same location, clicks cycle through them in descending order of size.
% \keep{To address nested structures, where multiple structures occupy the same location, clicks cycle through them in descending order of size.}

\paragraph{Computing Over Repetitive Hierarchical Structure.}
Finally, Sam wants to create a shopping list of all the ingredients.
In Cellery, formulas may resolve to whole structures---not just single values---and \syslists{} allow broadcast access to the members' properties. For example, writing the formula \formulatext{=Recipes.Recipe~Name} in a new grid cell will make a computed \syslist{}, with the recipe names spilling to adjacent grid cells. In a new grid cell below her recipes, Sam writes a formula leveraging her nested structures: \\[0.2em]
\centerline{\small\formulatext{=UNIQUE(CONCATENATE(Recipes.Ingredients[Ingredient]))}}\\[0.2em] This formula gets the \systemtext{Ingredients} \systable{} of each \sysblock{} in \systemtext{Recipes}, accesses the \systemtext{Ingredient} column of each, concatenates them into a single \syslist{}, and then deduplicates the \syslist{}.
The result is shown at the bottom of \Cref{teaser}.
Since formulas recompute when referenced structures change, adding recipes or ingredients automatically updates Sam's grocery list.
Sam is ready for her trip to the store!
%In comparison, traditional spreadsheet systems cannot express references to sets of nested structures, resulting in formulas that contain fragile lists of references.

% \jy{if space, talk about why this example would be hard in Excel. It forces users to choose between a more natural representation with less expressive formulas and more manual effort, vs. needing to reshape and flatten it to fit the computational model}

\section{System Description}

\Cref{structurestable} displays Cellery's core structural features. In the following sections, we detail the design of our system.

\subsection{Structure Model}
\label{model}
% Describe model and primitives
% Don't forget to mention sheets
% Fundamentally, the model is a hierarchical tree of labeled structures, which may reference other structures via labelling or ordering
% "everything is a structure", shared interface for structures applies first-class properties consistently. Connect back to composability and graduality
% Structures have position, dimensions, and optional formula. Atoms have values while lists, tables, and blocks have children
% Two types of cell structures: Data and Label

Cellery's design centers on a hierarchy of named structures representing the underlying organization and content of the spreadsheet. Following \dptwo, this hierarchy is built by composing four structure types.  Each sheet in Cellery is a collection of optionally-named structures, each of which may contain other structures, forming a tree. Some structures are \textit{containers} that store other structures (internal tree nodes); others are \sysatoms{} that hold values (leaf nodes). Following \dpone, this design supports \textit{partial} structural commitment: users can choose to organize some values on the grid into containers while leaving others as loose cells. Cellery tracks the current grid location of each structure, updating that location as needed to account for changes to adjacent structures.
% \Cref{fig:structures} shows an example tree and how its structures may map to a grid.

% \begin{figure}
% \vspace{2in}
% \caption{TODO}
% \label{fig:structures}
% \jy{Figure showing a structurally-rich spreadsheet in Cellery and its underlying tree model}    
% \end{figure}

Cellery rests on the principle that \textit{everything on the grid is a structure}, including cells. While traditional spreadsheet systems center the grid by treating features like named ranges and tables as ranges of cells, Cellery's structure-first model reframes cells themselves as structures.
%\jy{this is crucial to the gradual nature of our system because it allows for valid partially-specified structure?}
All structures have shared properties: they are rectangular, have a position and dimensions, an optional name, and an optional formula. Structures also share many capabilities: they can be interacted with in similar ways, can be referenced by name in formulas and assigned a formula (\Cref{formulas}), and can be incorporated in templates (\Cref{templates}). This ensures that structures behave consistently: our intent is to allow users to transfer concepts across structures of different types.

\subsection{Structure Primitives}
\label{primitives}
%\jy{Framing wise there is an opportunity to group some features together that we are not taking advantage of: Atoms + labels -> Gap 1, Tables + hierarchical tables -> Gap 4, Blocks + templates -> Gap 3 + 5}
% Table summary of primitives with categorical axes: dimensionality, ordering, named children, type of relationship
In this section, we describe Cellery's four base structure types: the primitives that compose together in our structure model.

\paragraph{\sysAtoms{}.} An \sysatom{} is a structure holding a single value, such as a number or a string. From the user's perspective, they function similarly to cells in other spreadsheet systems: users can freely type inside squares on the grid to create them.\ \sysAtoms{} may contain either raw values, or they can be assigned a formula.\ \sysAtoms{} are the only structures that can contain raw values, so collections storing data must be created by composing \sysAtoms{} together with other structure types.\ \sysAtoms{} can also serve as structure labels.
%\mc{why do we need cells to be designated? Why can't the user just type in grid locations as usual?}

\paragraph{\sysLists{}.} \sysLists{} are one-dimensional, ordered collections of other structures.
%While they can be approximated in traditional spreadsheets with  single-column tables, \syslists{} represent ordered sequences, rather than records with attributes. 
%We treat  as a primitive to capture this distinction, as well as to 
%\sysLists{} enable composability and formula semantics that the rest of the system depends on.
They can be either horizontal or vertical, and can be rotated using the context menu. In the spirit of supporting top-down and bottom-up structure, \syslists{} (and other types of containers) can be created on top of existing Atoms, or Atoms can be added by editing grid locations within.
%\sysLists{} can also contain empty elements.
By default, \syslists{} contain Atoms; however, to support composability, the \textit{type} of a \syslist{} can be changed using the structure panel.\ \syslists{} can contain any structure type (e.g., a \syslist{} of \systables). However, direct children of \syslists{} must be homogeneously typed: creating a \syslist{} with differently-typed elements (e.g., a \syslist{} containing an Atom and a \systable) is disallowed. This prevents ambiguity in formula semantics: for instance, \formulatext{SUM(list1)} produces a single value when \systemtext{list1} contains Atoms, but the sum of a \syslist{} containing both Atoms and \sysblocks{} would be ill-typed. However, due to Cellery's lack of a more expressive type system, this restriction does not apply across descendants: a \syslist{} containing a \syslist{} of Atoms and a \syslist{} of \systables{}, for instance, is allowed.
%Note that the types we describe refer to types of structures, not types of values; a \syslist{} of both strings and numbers, for instance, is permitted. Changing the type of a \syslist{} clears its values.
%\jy{is there a better way to describe this typing relationship?} \mc{I'd say the constraint is shallow. But then we need to say why this is.}
% Multi level headers, drag header edge to add/remove levels, default col headers

\paragraph{\sysTables{}.} \sysTables{} are two-dimensional, ordered, optionally axis-labeled collections containing other structures.
In our content analysis, we observed tables with a variety of header configurations. As a result, Cellery's \systables{} can be labeled with both column and row headers. Like Excel, headers take up grid space, and their values serve as column/row labels. To prevent ambiguity when accessing data, each row-column intersection may contain only one structure.

% Rows and columns can be selected by clicking on the associated header.\mc{leave interaction for later}
%(or by clicking near the edge of a row/column without a header).
%Doing so displays the properties of the selected row/column in the structure panel.
Like \syslists{}, \systables{} contain Atoms by default. To further support nested structure, \systables{} are typed at the column level: a structure type can be assigned to a selected \systable{} column.
% To resolve ambiguity in cases where a typed row and column intersect, the most recently applied type takes priority.\mc{weird.}
\sysTable{} rows and columns can be assigned formulas in a similar way.

%\mc{Maybe make the main example include all the types of structures. Maybe leave templates for another figure.}
% Figure demonstrating hierarchical table creation

Hierarchical tables appeared in 40\% of spreadsheets during our content analysis. For example, header columns ``2025'' and ``2026'' might each have sub-headers ``Q1''--``Q4,'' representing four quarters per year. To represent this structure in Cellery, hierarchical \systables{} can be created using \textit{multi-level headers}. ``Levels'' can be added to or removed from headers by dragging the edge of the header area. In our content analysis, 66\% of spreadsheets incorporating hierarchical tables did so using merged headers: Cellery allows users to merge upper-level header cells by dragging their edges to create \textit{column and row groups}, which can be accessed in formulas.
%This interaction is identical to the way merged Atoms are produced in Cellery (\Cref{dm}).
% This can be used to produce hierarchical groupings of \systable{} rows and columns.

\paragraph{\sysBlocks{}.} 37\% of spreadsheets in our content analysis contained non-tabular groupings of cells. Cellery supports these structures with \sysblocks{}, which are unordered rectangular regions on the grid that may contain other structures. Unlike other structure types, \sysblocks{} may contain heterogeneously typed children, e.g., an Atom, a \syslist{}, and a \systable{}. A \sysblock{} places almost no restrictions on what can be placed inside and functions simply as a grouping of structures on the grid.\ \sysBlocks{} are analogous to \emph{records} in functional programming because they consist of named fields. Finally, \sysBlocks{} can created from or converted to a template (\Cref{templates}).
% Unlike object instances in systems like ClassSheets and Object Spreadsheets, \sysblocks{} impose no schema or type constraints on their children: they are groupings rather than typed objects.
%\jy{Should i explicitly connect back to our gap about nontabular groupings or is it obvious?} \mc{Yes, and also relate to objects in related systems like ClassSheets.}

\begin{table}[!tb]
\caption{Summary of Cellery's structure types, grounded with codes from our content analysis.}
\centering
\small
\begin{tabular}{@{}p{1.7cm}p{4cm}p{2cm}@{}}
\toprule
\textbf{Structure} & \textbf{Description} & \textbf{Code} \\
\midrule
\styleatom{\textbf{\sysAtom}}  & A single value on the grid. & N/A
\\
\hspace{0.4em} \syslabel{} & Atom naming another structure. & Label cells
\\
\stylelist{\textbf{\sysList}}  & 1D, ordered collection of structures. & 1D collection
\\
\styletable{\textbf{\sysTable}} & 2D, tabular collection of structures. & Table
\\
\hspace{0.4em} Hierarchical & Table with hierarchical headers. & Header hierarchy
\\
\styleblock{\textbf{\sysBlock}} & Unordered grouping of structures. & Non-tabular grp.
\\
\hspace{0.5em} Template & Reusable Block with default values. & Repeated pattern
\\
\bottomrule
\end{tabular}
\label{structurestable}
\end{table}

\subsection{Structure-Aware Formula Language}
\label{formulas}
% Referencing (by name). Structures given default names
% Derivation: formulas can be assigned to or return structures
% Formula result edit protection
% Data accesses (indexing and dot notation)
% Array logic: elementwise arithmetic
% Functions
% Aggregation function directionality thing?
% Shorthand for writing maps: broadcasting
% Data can still be referenced by location

% Make sure to connect structure types to model via data accesses: lists and tables reference children via index while objects reference children by name

At a high level, Cellery's formula language is similar to that of other spreadsheets, extended with features for referencing and extracting data from our structures. To remain accessible to traditional spreadsheet users, Cellery adopts an Excel-like syntax. However, because structures can change locations, formulas normally reference them by \emph{name}.
For example, the formula \formulatext{=SUM(list1)} references the elements of \systemtext{list1}.
Nevertheless, to maintain compatibility with traditional spreadsheet language features, cell values can also be referenced using geometric addresses: e.g., \gridloc{A1} or \gridloc{B1:B3}.
% However, geomeric references in Cellery formulas do not change after structures are moved, so named references are preferable \bh{presumably this is just an unimplemented feature rather than a design choice?} \jy{Yes :( }.

To support references to data inside structures, Cellery uses syntax and semantics inspired by programming languages like Python. Elements of \syslists{} can be accessed by index, e.g., \formulatext{list1[0]}.\ \sysTables{} can be accessed by column name, e.g.\ \formulatext{Ingredients[Ingredient]}.\ \sysTables{} can also be accessed by row index, e.g.\ \formulatext{Ingredients[0]}, or by row label, if provided.\ \sysTable{} column and row references return results in the form of a \syslist{}; one limitation of this approach is that rows produced in this way cannot be accessed by column name afterwards, as there is no dedicated structure type for individual \systable{} records. Lastly, structures inside of \sysblocks{} can be accessed using dot notation, similarly to object fields in Python: e.g., \formulatext{block1.struct1}.

Cellery's formula language supports commonly-used functions (\systemtext{SUM}, \systemtext{AVERAGE}, \systemtext{COUNT}, etc.). Functions may operate over Atoms, \syslists{}, and \systables{}. Some functions are polymorphic: e.g., the \systemtext{SUM} function produces a scalar Atom value when called on a \syslist{}, but produces a \syslist{} when called on a \systable{}. Aggregator functions like \systemtext{SUM} take an optional axis parameter (\systemtext{DOWN} or \systemtext{ACROSS}) when called on \systables; when omitted, the axis is inferred based on the direction of the target \syslist{}. A vertical \syslist{} automatically aggregates across the rows of a \systable{}, and a horizontal \syslist{} aggregates down the columns.

Formulas inside containers can leverage their container for scope. Formulas inside a \sysblock{} can reference other structures in the \sysblock{} without having to specify the container; formulas assigned to \systable{} columns can reference other columns in the same \systable{}. This also allows formulas to be duplicated across template instances, since each is evaluated in its own context (\Cref{templates}).

% When a formula returns an entire structure as output, it is appropriate to \emph{assign} the formula to a structure, which takes the form of the formula's output.
Formulas are capable of returning entire structures as output. Using the structure panel, users can select a structure (e.g., a \syslist) and enter a formula that determines its values. Formulas can also be assigned to \systable{} columns, which behave identically to \syslists{} in the formula language. Additionally, writing a formula returning a structure directly on the grid creates a structure of the formula's output type. For instance, \formulatext{=block1} outputs a live copy of the \sysblock{} onto the grid, including all structures nested inside. Formulas referencing structures recompute automatically when those structures change. However, structures \textit{computed} by formulas are read-only: if a user attempts to edit such a structure, Cellery emits a warning and severs the link between formula inputs and output.

References to row and column groups also evaluate to entire \systable{} structures, enabling aggregate computations. For example, the formula \formulatext{=AVERAGE(2025)} inside of a \systable{} where \systemtext{2025} is a group of four columns computes the average across all four columns. A formula with a column group as output, such as \formulatext{=table1[colGroup]}, returns a whole \systable{} with all of the columns in the group.

Cellery has a \textit{broadcasting} feature for aggregate operations across \syslists{} of \sysblocks{}. %In the grocery list example (\Cref{example}), Sam needed to extract a \systable{} column from each \sysblock{} in a \syslist---an operation that might require a map and a lambda function in a functional programming language, which may not be accessible to spreadsheet users. 
Inspired by functional programming, users can map operations across elements of \syslists{}. However, doing so in a traditional functional language might require writing a lambda function, which may not be accessible to spreadsheet users without programming experience.
In Cellery, dot notation on a \syslist{} of \sysblocks{} automatically distributes over its elements: \formulatext{=list1.struct}, where \systemtext{list1} contains \sysblocks{}, returns a \syslist{} with \systemtext{struct} from each \sysblock{}.
% If \systemtext{a} is missing on one of the nested Blocks, the corresponding element in the output List is empty. 

\subsection{Template System}
\label{templates}

% Templates can be assigned to lists and columns/rows

33\% of the spreadsheets in our content analysis incorporated repeated structural patterns. To capture these in Cellery, we implemented \textit{templates}: customizable abstractions over structures that enable pattern reuse. Templates are managed and created using the \textit{template panel} on the left side of the screen (visible in \Cref{teaser}). Creating a new template opens up the template editor (shown in \Cref{fig:example2}~\circleletter{K}), a separate, resizable grid with the same capabilities as the main interface. The template editor contains a template definition: a \sysblock{} whose structures serve as the pattern to be reused. Values inside a template definition serve as defaults, while empty structures act as placeholders for instance-specific data. Formulas and \syslabels{} inside a template are preserved in instances, enabling formula reuse. Once defined, a template can be instantiated by dragging it from the template panel onto the grid, producing a \sysblock{} that copies all structures from the definition. Instances can be edited individually; changes to the template definition do not propagate automatically to preserve edits users have made to individual instances. A ``Sync All Instances'' button in the template editor applies updates to all instances, which may overwrite existing data.

To support bottom-up structuring (avoiding premature commitment~\cite{blackwell_cognitive_2001}), a \sysblock{} can also be converted into a template via the context menu. During conversion, contained \syslists{} and \systable{} rows are cleared, since these structures may hold variable-length data that differs across instances; Atoms retain their values as defaults. Templates can serve as element types for \syslists{} and \systable{} columns: selecting a template from the element type dropdown configures the container so that each element follows the template's structure. The same can also be accomplished by dragging a template from the panel or an existing template instance into an empty \syslist{}.
%In the recipe book example (\Cref{example}), each \sysblock{} in the \systemtext{Recipes} \syslist{} contains the template's \systemtext{Ingredients} \systable{}, \systemtext{Steps} \syslist{}, and associated Atoms.
Adding new elements to such a \syslist{} automatically produces new template instances: in the example, pressing the ``+'' button produces a new, fully structured recipe. This integrates naturally with the formula language's broadcasting feature, which extracts data across template instances in a structured way.

\subsection{Implementation Details}

We implemented Cellery as a web application in TypeScript with React. Cellery can import and export spreadsheets as JSON. Excel files can also be imported as unstructured spreadsheets of \sysatoms.
% The implementation uses a layered architecture separating the structure model and its associated operations from the user interface layer, which renders Atoms as values on the grid and other structures as overlays.
% The model maintains the tree of structures, an R-tree spatial index for fast retrieval of structures based on grid location (e.g., for clicking on structures), a dependency graph for formula recalculation, and template definitions. Children of container structures are positioned based on their offset from the origin of the parent: this improves the performance of operations like structure moves.
% \bh{It's a hierarchical scene graph!}
% \mc{will it be available as open source?}

% For our prototype, all data is stored on the client side.
% Could talk more about architecture, but it's not really interesting or research related

\section{Evaluation}

We evaluate Cellery using an expressiveness study and a preliminary usability study.

\begin{figure}[!htb]
    \centering
    \includegraphics[width=0.5\textwidth]{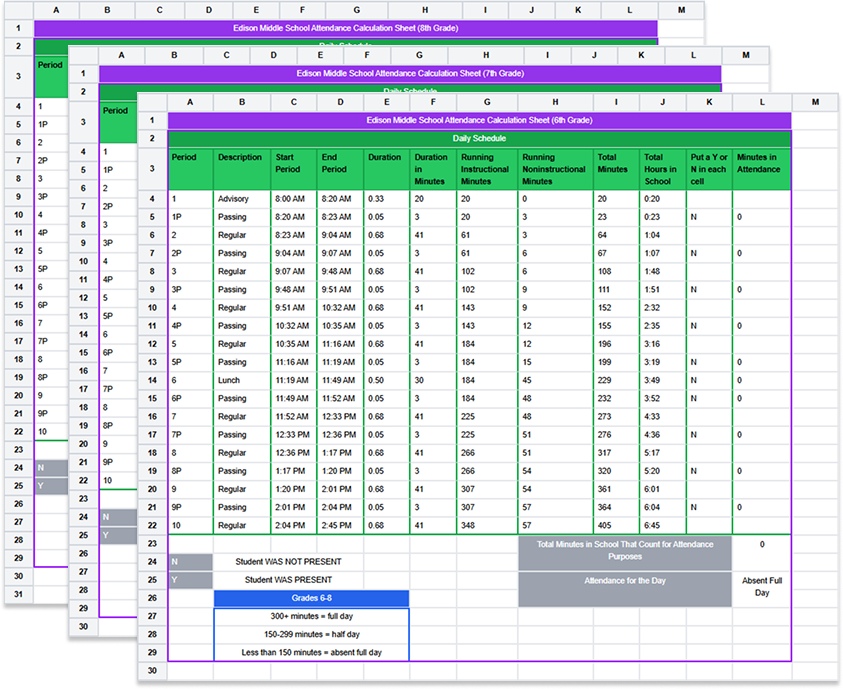}
    \caption{A spreadsheet, originally in Excel, reproduced in Cellery during our expressiveness study. The spreadsheet has three sheets with repeated structure, which in Cellery are linked together via a template. Of our five representational gaps, this spreadsheet expresses four: nested structure, labels, non-tabular groupings, and repeated patterns.}
    % Nested structure, labels on the grid, non-tabular groupings, repeated patterns
    \label{fig:reproduced}
    \Description[A spreadsheet reproduced in Cellery as part of the expressiveness study.]{A spreadsheet reproduced in Cellery as part of the expressiveness study. The figure displays multiple stacked spreadsheets representing different sheets in the same workbook. Each sheet has a similar structure with a table computing the attendance of a middle school grade: there are three sheets, one each for 6th, 7th, and 8th grade. Each sheet also contains labeled cells (e.g., one showing whether the student was absent or present for the day), and a List showing the number of minutes required for each of a full day, half day, and absent full day.
}
\end{figure}

\begin{table*}[!htb]
\centering
\small
\caption{Study participant demographics.}
\label{tab:participants}
\begin{tabular}{clllcll}
\toprule
\textbf{Participant \#} & \textbf{Spreadsheet Exp.} & \textbf{Programming Exp.} & \textbf{Occupation} & \textbf{Gender} & \textbf{Age} & \textbf{Education} \\
\midrule
1 & Intermediate & Novice       & Law student            & M & 25--34 & Bachelor's \\
2 & Intermediate & Novice       & Sustainability         & F & 18--24 & Bachelor's \\
3 & Intermediate & Expert       & Game developer         & M & 18--24 & Bachelor's \\
4 & Intermediate & Novice       & Film                   & M & 25--34 & Bachelor's \\
5 & Beginner     & Advanced     & Student (CS \& Finance)& M & 18--24 & High school \\
\bottomrule
\end{tabular}
\end{table*}

\subsection{Expressiveness Study}

We reproduced our original sample of 30 spreadsheets in Cellery to answer the research question: \textit{To what extent does Cellery address the representational gaps identified during our content analysis?}

\paragraph{Methods.} For each spreadsheet, the first author examined the original file and (1) imported it into Cellery using Cellery's Excel import feature; (2) constructed the spreadsheet's organizational structure in Cellery, creating structures corresponding to those observed during the content analysis. We prioritized structural fidelity over exact appearance, focusing on capturing the underlying organization of each spreadsheet in Cellery's model. Visual layout was preserved wherever possible; we noted cases needing alteration.

Because the representational gaps identified in \Cref{gaps} represent the structural limitations that Cellery aims to address, we focus our evaluation around these five organizational features. For each reproduction, we recorded whether each instance of each gap type was successfully expressible in Cellery. To document Cellery's limitations, we noted any adjustments needed to express each structure in the system, alongside any types of structure that Cellery's model was not capable of representing.

% Compare visual layout (any visual changes needed to be made?) and logical structure (anything that couldn't be represented?)

\paragraph{Results.} Every instance of each of the five representational gap types from our content analysis was expressible in Cellery, though some layout adjustments were required, which we now describe.
\Cref{fig:reproduced} displays an example of a reproduced spreadsheet that expresses four of the five gap types.
% Next, we describe instances of structure types that were not representable.

\textbf{Adjustments:} Twelve spreadsheets required minor adjustments. Since row and column hierarchies must be specified in Cellery using merged, multi-level headers, any tables whose hierarchy was indicated by styling rather than merging (7 spreadsheets) required adjustment. This typically involved adding a header level to the table, relocating the subordinate rows/columns to a different header level, and merging to produce row/column groups. Tables with rows or columns wider or taller than a single cell but containing single values (4 spreadsheets) required compaction, since manually resizing \systable{} rows/columns is not implemented in Cellery. Additionally, one spreadsheet used cell background color to indicate boolean values (true/false encoded by gray/white color); lacking styling features, we converted these to use TRUE/FALSE.

\textbf{Non-representable structures:} Three spreadsheets contained structural features that were not representable in Cellery. One spreadsheet contained row headers on the \textit{right} side of the table, while Cellery only supports headers on the left-hand side. Another spreadsheet included a flowchart constructed from bold cell borders, and the third contained a diagram of a GUI layout. In those documents, the spreadsheet is used as a canvas for drawing rather than a container for grid-structured data; this falls outside our scope.

\paragraph{Limitations.} We built Cellery to address the gaps identified during our content analysis: their representability follows from the system's design, and so the adjustments and non-representable structures characterizing the limits of our model should be viewed as the substantive findings of this expressiveness study. We conducted the study on the same sample on which we based our system: our results may not generalize to unseen spreadsheets. Additionally, due to Cellery's limited formula language, our study does not evaluate whether \textit{formulas} from the sample are reproducible in Cellery, only types of structure. A deeper analysis with formulas could evaluate our language features and reveal other limitations.

\subsection{Preliminary User Study}

\begin{table*}[!tb]
  \centering
  \footnotesize
  \setlength{\tabcolsep}{4pt}
  \renewcommand{\arraystretch}{1.05}
  \newcolumntype{L}{>{\raggedright\arraybackslash}X}

  \newlength{\pnumwd}\setlength{\pnumwd}{2em}
  \newlength{\colwd}
  \setlength{\colwd}{\dimexpr(\textwidth - 12\tabcolsep - \pnumwd)/6\relax}
  \newcolumntype{P}{>{\centering\arraybackslash}m{\pnumwd}}
  \newcolumntype{L}{>{\raggedright\arraybackslash}m{\colwd}}

  \caption{Summary of participants' approaches to the user study tasks,
    categorized by the structures and formulas used. Unique features of
    Cellery are indicated in \textbf{bold}. Note that some traditional
    spreadsheets (e.g., Excel) support tables and formulas referencing table
    columns, so these features are not indicated as unique to Cellery.}
  \label{tab:studyresults}

  \begin{tabular}{@{}P LLLLLL@{}}
    \toprule
    & \multicolumn{2}{c}{\textbf{Task 1}}
    & \multicolumn{2}{c}{\textbf{Task 2}}
    & \multicolumn{2}{c}{\textbf{Task 3}} \\
    \cmidrule(lr){2-3} \cmidrule(lr){4-5} \cmidrule(lr){6-7}
    \textbf{P\#}
    & \multicolumn{1}{c}{\textbf{Structures}} & \multicolumn{1}{c}{\textbf{Formulas}}
    & \multicolumn{1}{c}{\textbf{Structures}} & \multicolumn{1}{c}{\textbf{Formulas}}
    & \multicolumn{1}{c}{\textbf{Structures}} & \multicolumn{1}{c}{\textbf{Formulas}} \\
    \midrule

    1
    & Flat \styletable{Table} of \styleatom{Atoms} 
    & References to \styletable{Table} columns; geometric references
    & \textbf{\styleblock{Blocks} containing hierarchical \styletable{Tables}}
    & \textbf{References to hierarchical \styletable{Table} columns}; geometric references
    & \multicolumn{2}{c}{Did not complete} \\

    2
    & \textbf{\styletable{Table} containing column of \stylelist{Lists}}
    & \textbf{References to \stylelist{Lists} in \styletable{Tables}}
    & \textbf{Templated \styleblock{Blocks}, containing \styletable{Tables}, containing columns of \stylelist{Lists}}
    & \textbf{References to \stylelist{Lists} in \styletable{Tables}}; references to \styletable{Table} columns; geometric references
    & \multicolumn{2}{c}{Did not complete} \\

    3
    & \textbf{\styletable{Table} containing column of \stylelist{Lists}}
    & \textbf{References to \stylelist{Lists} in \styletable{Tables}}
    & \textbf{Templated \styleblock{Blocks}, containing \styletable{Tables}, containing columns of \stylelist{Lists}; Label \styleatom{Atoms}}
    & \textbf{References to \stylelist{Lists} in \styletable{Tables}; references to \styleatom{Atoms} in \styleblock{Blocks}}; references to \styletable{Table} columns
    & \textbf{\stylelist{List} of templated \styleblock{Blocks} containing \styleblock{Blocks}, \styletable{Tables}, \stylelist{Lists}, and \styleatom{Atoms}}
    & \textbf{References to \styletable{Table} columns in \styleblock{Blocks}; references to \styleatom{Atoms} in \styleblock{Blocks}} \\

    4
    & \textbf{\styletable{Table} containing column of \stylelist{Lists}}
    & \textbf{References to \stylelist{Lists} in \styletable{Tables}}
    & \textbf{Templated \styleblock{Blocks} containing hierarchical \styletable{Tables}}
    & References to \styletable{Table} columns; geometric references
    & \multicolumn{2}{c}{Did not complete} \\

    5
    & \textbf{\styletable{Table} containing column of \stylelist{Lists}}
    & \textbf{References to \stylelist{Lists} in \styletable{Tables}}
    & \textbf{\styleblock{Blocks} containing hierarchical \styletable{Tables}}
    & References to \styletable{Table} columns; geometric references
    & \multicolumn{2}{c}{Did not complete} \\

    \bottomrule
  \end{tabular}
\end{table*}

We conducted a think-aloud usability study asking: \textit{In tasks involving spreadsheet authoring and modification, what kinds of structure do users employ with Cellery, and what difficulties do they encounter when applying structural concepts?}

\paragraph{Participants.} We recruited a convenience sample of five participants (four men, one woman), all of whom had some spreadsheet experience.\ \Cref{tab:participants} displays participant demographics.
%Four of the six participants were recent graduates, one was a graduate student, and one was an undergraduate.
Participants were compensated with a \$30 gift card for 90 minutes of their time.

\paragraph{Protocol.} Participants watched a 16-minute video tutorial introducing Cellery. They then completed three spreadsheet tasks. We asked participants to think aloud and  allowed them to ask the researcher questions about the system. We captured recordings of the screen and took notes about each participant's approach.
%Following the session, participants completed a survey with a System Usability Scale (SUS) questionnaire~\cite{brooke_sus_1996}, as well as a question asking for feedback on the system.

\paragraph{Tasks.} To elicit gradual structure from participants, we separated tasks into parts that incrementally added complexity to the spreadsheet. The first task involved creating a spreadsheet recording the topic, names of attendees, and attendance count at a recurring meeting, then computing the average number of attendees across all meetings and a list of topics covered. The second task, adapted from a spreadsheet task used by Liang et al.~\cite{liang_tabletalk_2025}, involved computing grades for a class of students using a weighted average of assignment scores, and then extending the spreadsheet to compute the average grade of multiple classes. The third task involved producing a spreadsheet similar to our example in \Cref{teaser} and \Cref{example}, with a series of recipes containing heterogeneously-structured data and computed ingredient counts based on a scale factor. Full task descriptions can be found in the appendix.
%At the end of the task, participants were asked to produce a deduplicated list of all ingredients (the formula described in \Cref{example}).

\paragraph{Results.} We synthesized observations across our session notes to identify strategies employed by users and difficulties they faced.\ \Cref{tab:studyresults} displays a summary of participants' approaches to the tasks, categorized by types of structure and formulas used. Participants generally used Cellery's unique features during the tasks, and were able to gradually structure their spreadsheets with various compositions of Cellery's primitives. Four of the participants used a nested \systable{} of \syslists{} to complete Task 1. All four designed table column formulas that automatically adjusted to the length of the nested lists. The remaining participant used a flat table, but expressed that there was likely a ``more efficient way'' to perform the task. On Task 2, two participants again used a \systable{} of \syslists; three created hierarchical tables. Three participants wrote formulas referencing hierarchical columns or nested \syslists{} in \systables{}, while two did not. Notably, one participant created hierarchical tables but used Excel-style geometric references to write formulas instead of using the hierarchy, perhaps attributable to discomfort with the formula language; another participant used primarily regular, non-hierarchical table column references. Additionally, to create repeated tabular structures during Task 2, three participants used templates. However, to duplicate a formula, two participants created instances first, then manually changed each instance to add the formula (instead of editing the template and then instantiating it); some participants may underestimate the capabilities of templates. Four participants did not complete Task 3 due to the limited duration of the study. The one participant who did complete the task used repeated templates to do so, but struggled to access data from within nested \sysblocks{}, suggesting that the scoping behavior of \sysblocks{} may introduce difficulties for users.

% Finally, the study uncovered a number of usability issues. Participants experienced issues with some interactions: for instance, they often lost track of which structure they had selected, leading them to assign formulas to the incorrect structure. 

\paragraph{Limitations.} Some participants encountered bugs, e.g., with formula recomputation; difficulties with these are excluded from our analysis. The short duration of the study is another limitation, as 4/5 participants did not have time to complete Task 3. Due to the think-aloud and open-ended nature of the study, we do not evaluate task completion timing or correctness. Our study was conducted with a relatively small sample of five participants and without comparison to traditional spreadsheets; further investigation is needed to more definitively assess Cellery's usability and utility.

\section{Discussion and Future Work}
\label{discussion}

Following our content analysis, system design, and evaluation, we draw takeaways about the limits and implications of our system.

\paragraph{Future work on the formula language.} From our two-part evaluation, we find that Cellery's composable model is able to represent types of structure that traditional spreadsheets leave implicit, and moreover that users are capable of incorporating these structures during spreadsheet tasks. However, for the tasks that we gave our participants, we also found that the primary opportunities for improvement lie not in \textit{expressing} structure, but in \textit{leveraging} it. Most participants were able to define rich compositions of structures in Cellery; while the majority of participants for each task were also able to write structural formulas, some participants struggled and fell back on Excel-style geometric references. Similarly, templates had mixed adoption. This points to a distinction between \textit{structural expressiveness} and \textit{structural formula literacy}: Cellery succeeds in its primary goal of making structure more expressive, but users may need more support for writing formulas and managing abstraction in order to effectively utilize richer structure. Future work could investigate how to scaffold the formula authoring experience, especially for spreadsheet users who lack experience programming with structured data. For instance, there may be opportunities for users to leverage structure \textit{without} writing textual formulas: e.g., using direct-manipulation operations.

Our content analysis also did not consider the structural role of formulas, nor did our expressiveness study reproduce formulas; further design and evaluation of the formula language is an important opportunity for future research. For instance, the formula language could be improved with a more expressive type system over references to Cellery's structures. A formula-focused corpus analysis and user studies investigating the human factors of the language could better evaluate whether formulas in Cellery can leverage structure effectively.

\paragraph{Future work on the structural model.} The Cellery model is not the only possible model for gradually structuring spreadsheets. Future work could explore alternative data models: for instance, \systables{} could be represented as \syslists{} of Records. The experience of structuring spreadsheets could also be further explored: structures in Cellery must be manually created, but future work could investigate how to automatically infer our structures from spreadsheets.

\paragraph{Implications of gradual and composable structure.} The problem of users working with implicit structure is not unique to spreadsheets. Other substrates, such as document editors and canvas-based design tools, could support richer computation over and interaction with user-defined structure if it were made explicit.

\section{Conclusion}

Cellery shows how the implicit structure of spreadsheets can be made explicit using a composable model of structure. Guided by a content analysis of 30 spreadsheets, as well as our principles of composability and graduality, we designed a model consisting of four primitives---Atoms, \syslists{}, \systables{}, and \sysblocks---and instantiated it in a system, Cellery, that leverages structure in formulas, interactions, and reusable templates. We evaluated our system by reproducing our sample of 30 spreadsheets, as well as conducting a preliminary user study. By reimagining the spreadsheet model without replacing the grid, our work shows that expressive structure in spreadsheets can be achieved without sacrificing flexibility.

\begin{acks}

We thank the anonymous reviewers for their thoughtful feedback, as well as our user study participants for their time and insights.
We also thank Joshua Horowitz, Devamardeep Hayatpur, the SALAD Lab, and the Pacific Programming Interfaces Confab for many valuable discussions that helped shape this work.

\end{acks}

\balance{}

\bibliographystyle{ACM-Reference-Format}
\bibliography{cellery}

\clearpage
\onecolumn
\appendix
\section{Expressiveness Study Examples}
\begin{figure}[!htb]
  \centering
  \includegraphics[width=0.89\textwidth]{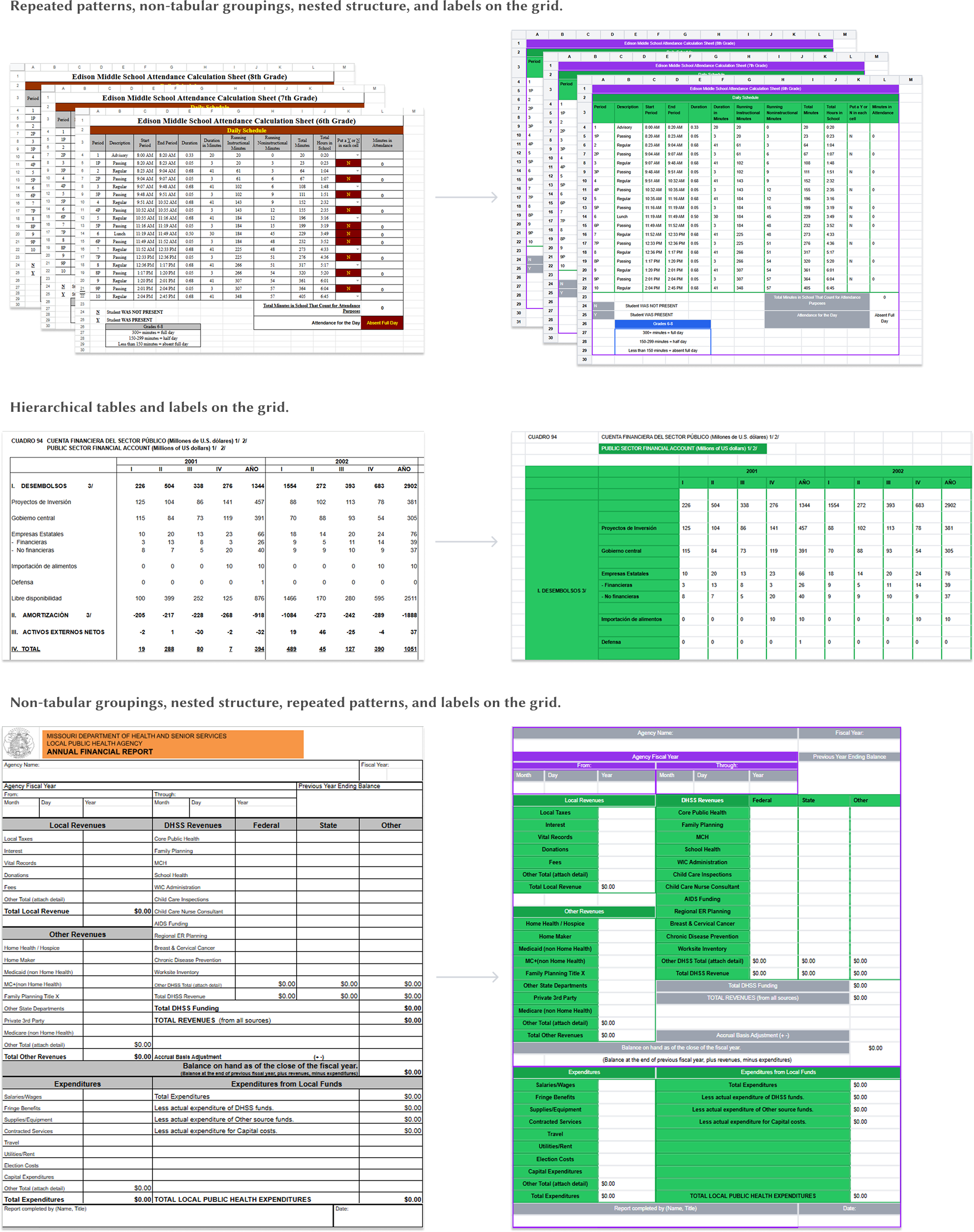}
  \caption{Examples of reproduced spreadsheets from our expressiveness study.}
  \label{fig:reproduction-study1}
  \Description[Examples from the expressiveness study showing 3 sets of Excel spreadsheets reproduced in Cellery.]{Examples from the expressiveness study showing 3 sets of Excel spreadsheets reproduced in Cellery. The first spreadsheet has a repeated pattern of 3 tables calculating attendance at a middle school across 3 different grades. Out of the representational gaps, this spreadsheet exhibits repeated patterns, non-tabular groupings, nested structure, and labels on the grid. The second spreadsheet shows a hierarchical financial table. It exhibits hierarchical tables and labels on the grid. The third spreadsheet shows an annual financial report consisting of tables and cells arranged in a rectangular fashion. It exhibits non-tabular groupings, nested structure, repeated patterns, and labels on the grid.}
\end{figure}
%\FloatBarrier

\begin{figure}[!htb]
  \centering
  \includegraphics[width=0.89\textwidth]{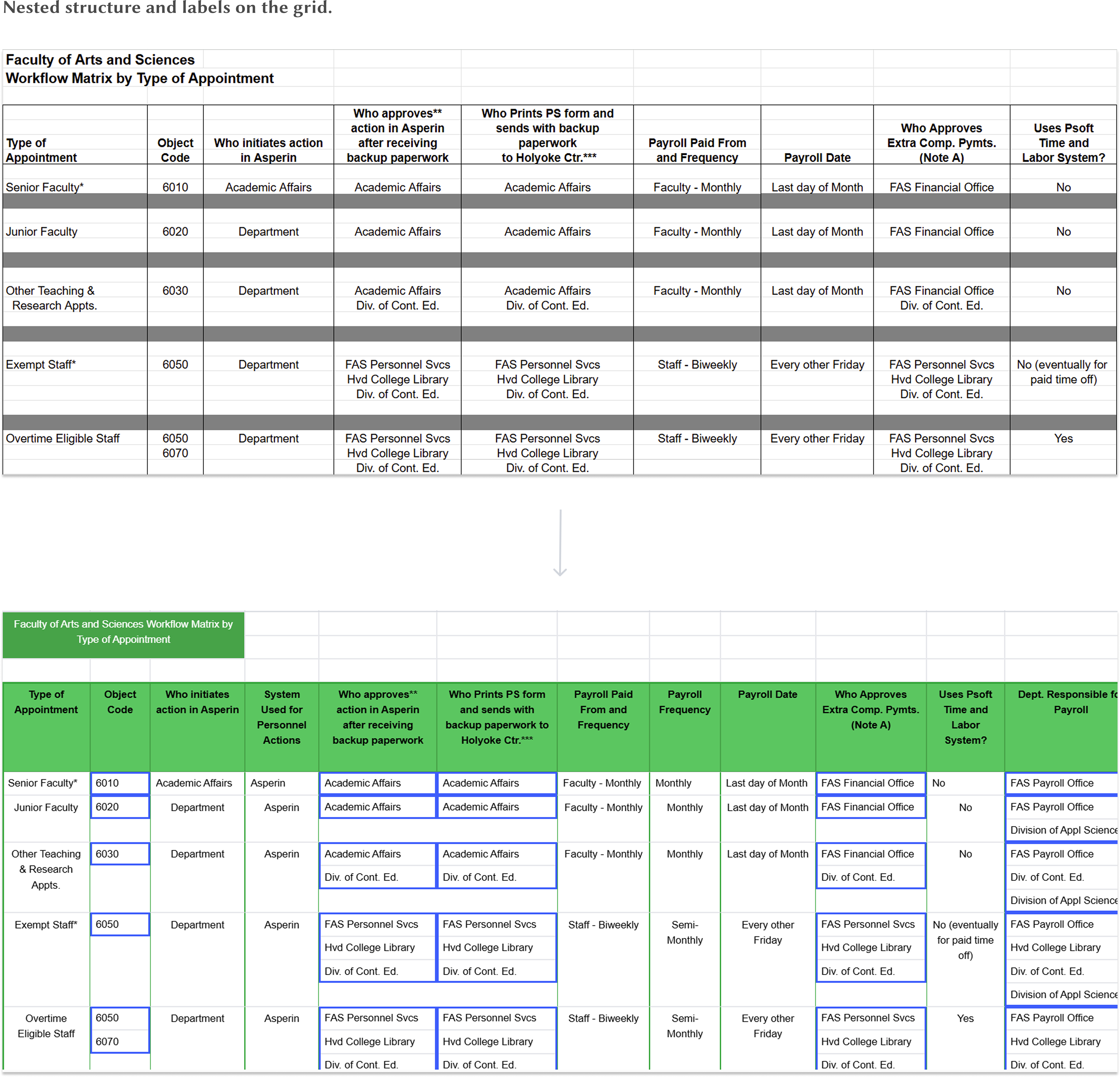}
  \caption{Example of a reproduced spreadsheet from our expressiveness study.}
  \label{fig:reproduction-study2}
  \Description[An example from the expressiveness study showing an Excel spreadsheet reproduced in Cellery.]{An example from the expressiveness study showing an Excel spreadsheet reproduced in Cellery. The spreadsheet has rows for different categories of faculty at an academic institution and columns showing e.g., what department approves their payments. Some row-column intersections contain multiple cell values. The spreadsheet exhibits nested structure and labels on the grid.
}
\end{figure}
%\FloatBarrier

\section{User Study Materials}
\label{studymaterials}

\subsection{Task 1}

\textbf{Part 1:}
You are a graduate student. Your research group has weekly meetings, and you want to record attendance at each week’s meetings this quarter. Create a spreadsheet that, for each meeting, records:
\begin{itemize}
    \item the week number
    \item the topic of the meeting
    \item the names of attendees
    \item the count of attendees
\end{itemize}
Make sure to design your spreadsheet so that new meetings and attendees can easily be added.

Data:
\begin{itemize}
    \item Week 1, ``Literature Review''; Attendees: Alice, Bob
    \item Week 2, ``Experiment Design''; Attendees: Alice, Carol, David
    \item Week 3, ``Related Work Discussion''; Attendees: Bob, Carol, Eve
\end{itemize}

\textbf{Part 2:}
Add two new meetings to the meeting record.

Data:
\begin{itemize}
    \item Week 4, ``Pilot Study Results''; Attendees: Alice, Bob, David, Grace
    \item Week 5, ``Writing Workshop''; Attendees: Alice, Frank
\end{itemize}

\textbf{Part 3:}
Using your spreadsheet, compute the following:
\begin{itemize}
    \item The average number of attendees across all meetings this quarter
    \item A list of the topics covered this quarter
\end{itemize}

\subsection{Task 2}

\textbf{Part 1:} 
You are a teacher who taught a course in geometry to high school students. Your job is to compute the overall grade of each student in the class, as well as the average grade across the whole class.
Each student's final grade is a weighted average of 3 types of assignments: homeworks, exams, and participation activity.
In your course, there were 2 homeworks, 2 exams, and 1 participation activity.
The assignments are worth 40\% of the grade, the exams are worth 50\% of the grade, and the participation activity is worth 10\% of the grade.
All grades should be expressed as percentages.
Create a spreadsheet computing the overall grade for each student in the class, as well as the average grade across the whole class. Make sure to design the spreadsheet so that new data can be added as necessary.

Data:
\begin{itemize}
    \item Maria: Homework 1, 90; Homework 2, 93; Exam 1, 94; Exam 2, 89; Participation, 95.
    \item David: Homework 1, 65; Homework 2, 70; Exam 1, 75; Exam 2, 73; Participation, 80.
    \item Aisha: Homework 1, 88; Homework 2, 84; Exam 1, 92; Exam 2, 87; Participation, 93.
\end{itemize}

\textbf{Part 2:}
You've finished grading assignments for a couple new students, and need to add them to the sheet. Add these students, making sure to compute their overall grade and update the class average accordingly.

Data:
\begin{itemize}
    \item Ethan: Homework 1, 78; Homework 2, 82; Exam 1, 88; Exam 2, 90; Participation, 87.
    \item Sofia: Homework 1, 95; Homework 2, 92; Exam 1, 91; Exam 2, 93; Participation, 98.
\end{itemize}

\textbf{Part 3:} 
It's a new semester, and now you have a different class of students. Also, this semester there is one more homework assignment (for a total of 3). Produce a new section in the spreadsheet for Semester 2 using the same grade calculation. As before, you should also compute the average for the new class.

Data:
\begin{itemize}
    \item Priya: Homework 1, 83; Homework 2, 79; Homework 3, 85; Exam 1, 76; Exam 2, 80; Participation, 88.
    \item Chi: Homework 1, 91; Homework 2, 89; Homework 3, 92; Exam 1, 90; Exam 2, 93; Participation, 97.
    \item Noah: Homework 1, 74; Homework 2, 77; Homework 3, 73; Exam 1, 70; Exam 2, 72; Participation, 85.
\end{itemize}

\textbf{Part 4:} 
After a couple years of teaching this class, you now have grades for 4 classes split into 2 years. You want to compute the average grade of each class, as well as the average grade across all classes in each year. Produce another section in the spreadsheet for Year 2, and calculate the average grade for each year.

Data:

Year 2, Semester 1:
\begin{itemize}
    \item Julio: Homework 1, 81; Homework 2, 85; Homework 3, 83; Exam 1, 78; Exam 2, 82; Participation, 88.
    \item Amir: Homework 1, 87; Homework 2, 89; Homework 3, 91; Exam 1, 90; Exam 2, 92; Participation, 94.
\end{itemize}

Year 2, Semester 2:
\begin{itemize}
    \item Chloe: Homework 1, 84; Homework 2, 88; Homework 3, 86; Exam 1, 89; Exam 2, 91; Participation, 93.
    \item Benji: Homework 1, 77; Homework 2, 79; Homework 3, 75; Exam 1, 72; Exam 2, 74; Participation, 85.
    \item Amelia: Homework 1, 93; Homework 2, 95; Homework 3, 94; Exam 1, 97; Exam 2, 98; Participation, 100.
\end{itemize}

\subsection{Task 3}
\textbf{Part 1:} 
You are planning to cook dinner for a party, and you want to create a spreadsheet with all of the recipes you’ll need. Start with one recipe. Your spreadsheet should include the following:
\begin{itemize}
    \item Recipe name
    \item Number of servings
    \item A table with ingredients and amount of each ingredient required
    \item A list of steps
\end{itemize}

Data:
\begin{itemize}
    \item Recipe name: Cookies
    \item Number of servings: 2
    \item Ingredients: 3 cups flour, 2 eggs, 1 cup butter, 1 cup sugar
    \item Steps:
1. Mix ingredients well.
2. Roll cookies onto a baking sheet.
3. Bake at 375F for 10 minutes.
4. Let cool for 5 minutes.
\end{itemize}

\textbf{Part 2:} 
You realized that you need to make more servings than your recipe provides. To solve this, you want to add a scale factor that you can multiply the servings and number of ingredients by (e.g.\ double the ingredients to make double the servings). Add the scale factor to your spreadsheet, along with a cell computing the new number of servings. Finally, add a column to the table of ingredients showing the new amount of each ingredient after applying the scale factor.

Data:
\begin{itemize}
    \item Scale factor: 2
\end{itemize}

\textbf{Part 3:} 
Now, you want to add more recipes to your recipe book. Add the following two recipes. They should follow the same pattern as the previous one: each recipe should have a recipe name, original number of servings, scale factor, new number of servings, ingredients table (with column applying the scale factor), and list of steps.

Data:

Recipe 1:
\begin{itemize}
    \item Recipe name: Braised Celery
    \item Number of servings: 4
    \item Scale factor: 3
    \item Ingredients: 1 bunch celery, 2 tablespoons butter, 0.5 cup chicken broth
    \item Steps: 1. Cut celery into 1-inch pieces. 2. Cook celery in butter over medium heat. 3. Add broth and simmer for 7 minutes.
\end{itemize}

Recipe 2:
\begin{itemize}
    \item Recipe name: Ants on a Log
    \item Number of servings: 10
    \item Scale factor: 2
    \item Ingredients: 1 bunch celery, 0.5 cup peanut butter, 0.25 cup raisins
    \item Steps:
1. Cut celery stalks in half. Spread peanut butter on top.
2. Arrange raisins on top of peanut butter.
\end{itemize}

\textbf{Part 4:} 
Finally, you want to create a shopping list of all the ingredients. Use a formula to get a list of the unique ingredient names across all recipes. Ideally, your formula should update automatically when new recipes or ingredients are added.

% \pagestyle{empty}
% \clearpage\circleletter{A}
% \clearpage\circleletter{B}
% \clearpage\circleletter{C}
% \clearpage\circleletter{D}
% \clearpage\circleletter{E}
% \clearpage\circleletter{F}
% \clearpage\circleletter{G}
% \clearpage\circleletter{H}
% \clearpage\circleletter{I}
% \clearpage\circleletter{J}
% \clearpage\circleletter{K}
% \clearpage\circleletter{L}
% \clearpage\circleletter{M}
% \clearpage\circleletter{N}
% \clearpage\circleletter{O}
% \clearpage\circleletter{P}
% \clearpage\circleletter{Q}
% \clearpage\circleletter{R}
% \clearpage\circleletter{S}
% \clearpage\circleletter{T}
% \clearpage\circleletter{U}
% \clearpage\circleletter{V}
% \clearpage\circleletter{W}
% \clearpage\circleletter{X}
% \clearpage\circleletter{Y}
% \clearpage\circleletter{Z}

\end{document}